\documentclass[12pt,english,onecolumn,ondesided]{ar-1col}
\usepackage[latin9]{inputenc}
\usepackage[letterpaper]{geometry}
\usepackage{babel}
\usepackage{array}
\usepackage{units}
\usepackage{amsmath}
\usepackage{amssymb}
\usepackage{graphicx}
\usepackage[authoryear]{natbib}
\usepackage[unicode=true,
 bookmarks=false,
 breaklinks=false,pdfborder={0 0 1},backref=section,colorlinks=false]
 {hyperref}

\makeatletter

\providecommand{\tabularnewline}{\\}

\usepackage{jour-abbrev}
\usepackage{bold-extra}
\usepackage{color}
\jname{Annu. Rev. Astron. Astrophys.} 
\jvol{55}
\jyear{2017}
\doi{10.1146/((please add article doi))}

\makeatother

\begin{document}

\global\long\def\Mo{M_{\odot}}
\global\long\def\Lo{L_{\odot}}
\global\long\def\Ro{R_{\odot}}
\global\long\def\Rs{R_{\star}}
\global\long\def\Ms{M_{\star}}
\global\long\def\ns{n_{\star}}
\global\long\def\vs{v_{\star}}
\global\long\def\Ns{N_{\star}}
\global\long\def\Mbh{M_{\bullet}}
\global\long\def\rbh{R_{\bullet}}
\global\long\def\rh{r_{h}}
\global\long\def\as{^{\prime\prime}}
\global\long\def\mag{^{\mathrm{m}}}
\global\long\def\SgrA{\mathrm{Sgr\,A^{\star}}}

\markboth{T. Alexander}{Stars near massive black holes}

\title{Stellar Dynamics and Stellar Phenomena Near A Massive Black Hole}
\author{Tal Alexander\affil{Department of Particle Physics and Astrophysics, Weizmann Institute of Science, 234 Herzl St, Rehovot, Israel 76100; email: tal.alexander@weizmann.ac.il}---\\
{\tt Author's original version. To appear in Annual Review of Astronomy and Astrophysics. See final published version in ARA\&A website:\\
{\footnotesize \url{www.annualreviews.org/doi/10.1146/annurev-astro-091916-055306}}}}
\begin{abstract}
Most galactic nuclei harbor a massive black hole (MBH), whose birth
and evolution are closely linked to those of its host galaxy. The
unique conditions near the MBH: high velocity and density in the steep
potential of a massive singular relativistic object, lead to unusual
modes of stellar birth, evolution, dynamics and death. A complex network
of dynamical mechanisms, operating on multiple timescales, deflect
stars to orbits that intercept the MBH. Such close encounters lead
to energetic interactions with observable signatures and consequences
for the evolution of the MBH and its stellar environment. Galactic
nuclei are astrophysical laboratories that test and challenge our
understanding of MBH formation, strong gravity, stellar dynamics,
and stellar physics. I review from a theoretical perspective the wide
range of stellar phenomena that occur near MBHs, focusing on the role
of stellar dynamics near an isolated MBH in a relaxed stellar cusp.

\end{abstract}
\maketitle \tableofcontents{}

\section{Introduction and outline}

\label{s:intro}

\subsection{Motivation}

Massive black holes (MBHs) with masses $\Mbh\sim10^{6}-10^{10}\,\Mo$
lie in the nuclei of most galaxies \citep[see review by][]{gra16}.
This conclusion follows from the cosmic statistics of active galactic
nuclei (AGN), which are powered by a MBH \citep[see review by][]{net15},
from the analysis of gas and stellar dynamics observed in well-resolved
galactic nuclei in the nearby universe \citep{fer+00,geb+00,kor+13},
from the rare edge-on maser emitting circumnuclear accretion disks
that allow precise dynamical measurements \citep{miy+95}, from comparing
the mean mass density in the cosmic UV background (assumed to be generated
by AGN) to the space density of galaxies \citep{sol82,yuq+02}, and
finally, from detailed studies of the faint accretion emission and
the stellar motions around $\SgrA$, the MBH closest to us in the
center of the Milky Way \citep[see reviews by][]{ale05,gen+10}, which
provide the strongest empirical evidence for the existence of a MBH. 

The ubiquity of MBHs and of their phases of massive outflows and huge
radiative output, which can outshine their host galaxy, place MBHs
at the crossroads of many phenomena and processes in the realms of
cosmology, astrophysics and strong gravity, both as active agents
and as probes.

MBHs in galactic nuclei are embedded in a dense stellar environment.
This is observed in the Milky Way and in nearby galaxies \citep[e.g.][]{lau+98,fri+16},
and is predicted theoretically for isolated galactic nuclei \citep{bah+77,you77b}.
Of the wide range of complex processes associated with MBHs and their
close environment, stellar phenomena are substantially more tractable
for analysis and modeling. This is because in many cases the stars
can be treated as point masses, which are affected only by gravity,
and are free from the micro-physical uncertainties of non-gravitational
forces that may dominate gas phase dynamics, such as radiation pressure,
magnetic fields and hydrodynamic turbulence. The very large ratio
between the mass of a star and that of the MBH allows to simplify
the problem even further by treating the stars as test masses on intermediate
distance and time scales where the stars are deep enough in the MBHs
potential so that star-star gravitational interactions are negligible,
but not so close to the MBH that tidal effects are important. In that
limit, where each star separately can be approximated as forming a
2-body system with the MBH, atomic physics can provide useful analogies
and tools for dealing with 3- and 4-body interactions \citep[see review by][]{ale05}.
For many purposes the much simpler Newtonian treatment of the dynamics
suffices, and General Relativity (GR) effects can be omitted. Even
when the finite stellar mass, size and lifespan can no longer be neglected,
and stellar evolution becomes relevant, the uncertainties in the analysis
and modeling can be reduced by drawing on the well-developed theoretical
understanding of stellar structure and evolution, and on the large
body of detailed stellar observations away from galactic nuclei, in
the low-density field, and in dense stellar clusters. 

The presence of a MBH in a high-density stellar cluster offers opportunities
to test theories of stellar structure and evolution. Dynamics near
MBHs with masses $<10^{8}\,\Mo$ are collisional \citep{ale11,bar+13}.
Stellar velocities close to the MBH exceed the escape velocity from
the stellar surface (i.e. the kinetic energy in a star-star collision
exceeds the stellar binding energy), and collision rates rise to a
level where many of the stars suffer strong encounters \citep[e.g.][]{ale+01a}.
Likewise, tidal encounters with the MBH can substantially perturb
the internal stellar structure \citep{ale+01b,man+13}. Such galactic
nuclei are effectively ``stellar colliders'' that probe the physics
of stellar interiors by smashing stars energetically against each
other and against the MBH. 

\subsection{MBHs and their stellar hosts}

\label{ss:host}

The MBH dominates the gravitational potential inside its radius of
influence, $\rh$, which can be estimated by equating the potential
of the MBH, were it not surrounded by stars, to the potential of the
galactic nucleus, in the absence of a MBH, $-G\Mbh/\rh=\phi_{\star}(\rh)$.
When $\phi_{\star}$ is approximated as a singular isothermal distribution
(stellar mass density $\rho_{\star}=\sigma^{2}/2\pi Gr^{2}$), whose
potential has a constant velocity dispersion $\sigma$, the radius
of influence is $\rh=G\Mbh/\sigma^{2}$, and the total mass in stars
within $\rh$ is ${\cal O}(\Mbh)$. This is a useful approximation
for galactic nuclei, since typically, stellar dynamical processes
drive the stellar distribution near an isolated MBH to a power-law
cusp\footnote{This is quite unlike the case of a cluster with a dense core of mass
$M_{c}$, size $R_{c}$, dispersion $\sigma_{c}^{2}\sim GM_{c}/R_{c}$,
and a (hypothetical) central intermediate mass black hole (IMBH) with
$\Mbh\ll M_{c}$. There the total mass in stars within $\rh$ is $\propto(\Mbh/M_{c})^{2}\Mbh\ll\Mbh$
\citep[e.g.][]{las+11}. }, not much different from an isothermal sphere (Sec. \ref{s:dynamics}). 

There are strong empirical correlations between the masses of MBHs
and the spheroidal component of their host galaxies (the bulge in
the case of spirals, or the entire galaxy in the case of ellipticals).
The stellar velocity dispersion $\sigma$ in the bulge (on length
scale $\gg\rh$) correlates with the MBH mass as $\Mbh=M_{0}(\sigma/\sigma_{0})^{\beta}$,
with $\beta\sim4-5$ \citep{fer+00,geb+00,mcc+11} (the ``$\Mbh/\sigma$
relation''). The MBH's mass scales with the bulge's mass $M_{b}$
as $\Mbh=\epsilon_{\bullet}M_{b}$, with $\epsilon_{\bullet}\sim(1-2)\times10^{-3}$
\citep{mag+98,hae+04}. These correlations with properties on scales
well beyond the MBH's direct dynamical reach suggest a fundamental
evolutionary link between MBHs and their host galaxies, which may
be due to feedback by the powerful outflows in the MBH's AGN phase,
but whose exact nature is still unclear \citep[see review by][]{kin+15}.

The presence of a dense central stellar cluster in a galactic nucleus
does not necessarily imply the presence of an MBH. Nuclear stellar
clusters with no detectable MBHs are observed in lower mass galaxies
($\lesssim10^{10}\,\Mo$), while the nucleus of very massive galaxies
($\gtrsim10^{12}\,\Mo$) is completely dominated by the MBH. MBHs
and nuclear stellar clusters often coexist in galaxies in the intermediate
mass range \citep{geo+16}. 

Conversely, the absence of a dense central stellar cusp\footnote{A cusp is a density distribution that formally diverges at $r\to0$,
e.g. a power-law $n_{\star}\propto r^{-\alpha}$ with $\alpha>0$.
A cored density distribution flattens to a finite central density
inside the core radius.} does not imply the absence of a MBH. Massive galaxies tend to have
cored nuclei. The extrapolation back to the origin of the decreasing
density profile outside the core supports the idea that ${\cal O}(\Mbh)$
of stellar mass was removed from a previously existing cusp \citep{kor+09,tho+14}.
A natural explanation is that past galactic mergers led to the formation
of a tight MBH binary, which decayed dynamically and coalesced to
form the present MBH by ejecting a stellar mass comparable to its
own from the nucleus, leaving behind a lower-density stellar core
\citep{beg+80,mil+02}.

AGN statistics imply that MBHs exist in all galaxy types \citep{kau+03}.
Some are found at high redshifts of $z\gtrsim7$, when the universe
was only $\lesssim5$\% of its present age, and yet they appear to
be already very massive then, with $\Mbh\gtrsim10^{9}\,\Mo$ \citep[e.g.][]{mor+11}.
This requires a very efficient formation process that allows MBHs
to grow very fast, very soon after the Big Bang. Early stellar populations
are implicated in some of the proposed scenarios for creating or rapidly
growing a massive BH seed that can jump-start the relatively slow
process of growth by disk accretion \citep{vol12}: Rapid stellar
collisions of low metallicity stars in very dense clusters can lead
to the runway formation of a very massive star by mergers, which will
retain most of its mass as it collapses directly into a black hole
(BH) \citep{omu+08,dev+09,dav+11}, or a stellar BH that is launched
into a phase of supra-exponential accretion from the dense cold inter-stellar
medium of high-$z$ proto-clusters by the random gravitational perturbations
of the other stars \citep{ale+14b}. 

Intermediate mass BHs (IMBHs), with masses between the ${\cal O}(10\,\Mo)$
scale of stellar BHs and the ${\cal O}(10^{6}\,\Mo)$ scale of the
lightest detected MBHs, are expected to exist in dense stellar clusters.
This is based on an extrapolation of the $\Mbh/\sigma$ relation to
low masses, and it also follows from various formation scenarios,
which either allow some IMBHs to form late, so they would not have
had time to grow further, or else their formation requires special
conditions, whose chance absence can stall the IMBH's growth. The
continuing lack of firm evidence for IMBHs remains puzzling \citep[see review by][]{mil+04b}. 

\subsection{Unique properties inside the MBH radius of influence}

\label{ss:unique}

The singular character of MBHs and the extreme properties of their
close environment make these systems into unique physical laboratories.

\paragraph*{A singular relativistic potential}

\label{p:RelPotential}

The MBH's deep gravitational potential dominates over stellar gravity
well inside $\rh$, and imposes on the system its spherical symmetry
(for a non-spinning Schwarzschild MBH) or axial symmetry (for a spinning
Kerr MBH). Because the dynamical effects of the MBH spin fall rapidly
with distance (above $r\gtrsim10r_{g},$ where $r_{g}=G\Mbh/c^{2}$
is the MBH's gravitational radius), the potential over $r_{g}\ll r<\rh$
is to leading order spherical irrespective of the MBH spin. This high
degree of approximate symmetry limits orbital evolution around the
MBH to nearly fixed Keplerian ellipses on nearly fixed planes. Poisson
fluctuations away from spherical symmetry, due to the finite number
of stars in the system, result in residual forces that exert coherent
(``resonant'') torques on the orbits, leading to rapid angular momentum
evolution (\citet{rau+96}; Sec. \ref{ss:RR}). 

Secular relativistic dynamical effects, such as the advance of the
angle of periastron (Schwarzschild precession), Lense-Thirring precession
of the orbital plane around a spinning MBH, Bardeen-Petterson torques
on an accretion disk \citep{bar+75}, or orbital decay by the emission
of gravitational waves (GWs), all fall rapidly with distance, but
can nevertheless have substantial effect over time even for mildly
relativistic orbits \citep[e.g.][]{lev+03}. The strong coherent Newtonian
torques and the relativistic effects are coupled. On the one hand,
the coherent torques compete with the more subtle relativistic effects,
and make tests of strong gravity very challenging \citep{mer+10}.
On the other hand, fast relativistic precession can suppress the coherent
torquing by adiabatic invariance, and decouple very relativistic orbits
from the perturbing background \citep{mer+11,bar+14}. In particular,
this effect allows stellar BHs to inspiral unperturbed into MBHs by
the emission of GWs (\citealt{hop+06a,bar+16}; Sec. \ref{ss:relLC}). 

In marked contrast to the shallower potentials of dense clusters,
the deep singular potential of the MBH allows the retention of a dense
concentration of stellar BHs \citep{pre+10}, which accumulate very
close to the center due to mass segregation (Sec. \ref{sss:SScusp}),
and play an important role in the acceleration of dynamical processes
there. 

\paragraph*{A strong tidal field}

\label{p:TidalField}

The diverging potential of the MBH allows the tidal force on any bound
object (a star, a binary, a gas cloud) with mass $\Ms$ and size $\Rs$,
$F_{t}\sim G\Mbh\Ms\Rs/r^{3}$, to exceed its self-gravity $F_{s}\sim G\Ms^{2}/\Rs^{2}$
if the distance $r$ from the MBH is small enough, $r<r_{t}\simeq(\Mbh/\Ms)^{1/3}\Rs$.
The object is then tidally disrupted and its unbound fragments fly
off on ballistic orbits. The tidal disruption condition can be stated
alternatively in terms of the distance where the star's mean density
falls below the density the MBH would have if its mass were spread
over $r_{t}$, $\rho_{\star}=\Ms/\Rs^{3}\le\Mbh/r_{t}^{3}$, or in
terms of the distance where the crossing time of the tidal disruption
zone falls below the star's dynamical time $\sqrt{r_{t}^{3}/G\Mbh}\le\sqrt{\Rs^{3}/G\Ms}=t_{\star}$.

The only limit on disruption by a MBH is set by the size of the event
horizon $\rbh=x_{\bullet}r_{g}$ ($x_{\bullet}=2$ for a non-spinning
BH; generally $1\le x_{\bullet}\le9$, with $x_{\bullet}=1$ for an
equatorial co-rotating orbit around a maximally spinning Kerr BH,
and $x_{\bullet}=9$ for an equatorial counter-rotating orbit). When
$r_{t}<\rbh$, the disruption is hidden and confined inside the event
horizon, and a distant observer sees a direct plunge. Since $r_{t}/\rbh\propto\rho_{\star}^{-1/3}x_{\bullet}^{-1}\Mbh^{-2/3}$,
there exists for a star of given mean density $\rho_{\star}$ a maximal
MBH mass that allows tidal disruption outside of the event horizon;
the higher the MBH spin, the larger is $\max\Mbh$; for Solar type
stars and a Schwarzschild MBH it is $\max\Mbh\sim10^{8}\,\Mo$ \citep[see review by][]{ale12}.
The lower the MBH mass, the denser the stars it can disrupt outside
$\rbh$. An IMBH can tidally disrupt even a white dwarf (WD), \citep{lum+89}.
Conversely, very dense low mass main sequence (MS) stars ($\rho_{\star}\propto\Ms^{-11/4}$
for $\Ms\lesssim1.5\,\Mo$, \citealt{sch+92a}), can survive the MBH
tidal field and inspiral by the emission of GWs \citep{fre03}.

Tidal disruption events (TDEs) by MBHs are expected to lead to short
accretion flares, which can signal the presence of an otherwise quiescent
MBH \citep{ree88}, and probe both the accretion physics and the dynamics
leading to the TDE (Sec. \ref{s:tide}). The number of observed candidate
TDEs is fast growing, and it is becoming clear that this class of
variable sources has a much richer phenomenology than captured by
the initial simplified models \citep{kom15}. 

\paragraph*{A high stellar density}

\label{p:HighDensity}

The stellar density distribution around an isolated MBH, well inside
$\rh$, is expected to rise sharply as an approximate power-law cusp
$n_{\star}\propto r^{-\alpha}$ (Sec. \ref{sss:SScusp}). This occurs
whether the system is formed adiabatically on a timescale shorter
that the 2-body relaxation time, in which case the power-law index
depends on the initial conditions, for example $\alpha=3/2$ when
the surrounding stellar system is isothermal \citep{you77b}, or whether
the system is dynamically relaxed, in which case $\alpha=7/4-5/2$,
depending on the stellar mass function \citep{bah+77,ale+09,kes+09,pre+10}.
For lighter MBHs, the mean stellar density inside $r_{h}$ can be
as high as that of the cores of the densest globular clusters (Sec.
\ref{ss:NR}); the local density rises rapidly towards the MBH, and
can reach densities that are a few orders of magnitude higher than
anywhere else in the universe. Since the rate of collisions per star
scales with $n_{\star}$, destructive stellar collisions \citep[e.g.][]{ale99}
and strong non-destructive tidal and grazing stellar encounters \citep{ale+01a}
become dynamically important near the MBH, and affect the stellar
population there.

The high density cusp can be destroyed, or prevented from forming
by external perturbations, such as galactic mergers that lead to the
formation of a MBH binary, which slings out the stars inside its orbit
as it decays \citep{beg+80,mil+01,mer+05}.

\paragraph*{High orbital velocities}

\label{p:HighVel}

The singular potential of the MBH allows stars to approach the MBH
up to the tidal disruption radius, and reach velocities of up to $v\sim\sqrt{2G\Mbh/r_{t}}\sim\sqrt{2}(\Mbh/\Ms)^{1/3}\vs\gg\vs$,
where $\vs=\sqrt{G\Ms/\Rs}$ is the break-up rotational velocity of
the star. For example, a Solar type star near the MBH of the Milky
Way can reach $v(r_{t})=\sqrt{2r_{g}/r_{t}}\gtrsim0.3c$ (Sec. \ref{ss:SgrA}).
These extreme velocities have several implications. Stellar encounters
at $v\gg\vs$ are too fast for efficient exchange of energy or angular
momentum at impact parameter larger than $\Rs$ (cf Eq. \ref{e:TNR}),
and therefore 2-body relaxation is suppressed inside the collision
radius, $r_{\mathrm{coll}},$ where the velocity dispersion $\sigma^{2}\sim G\Mbh/r$
satisfies $\sigma(r_{\mathrm{coll}})\sim\vs$ (Sec. \ref{ss:NR}).
Direct stellar collisions are destructive, since the kinetic orbital
energy at infinity of the colliding stars exceeds the stellar binding
energy (Sec.\ref{ss:2StarColl}). Likewise, binaries cannot survive
3-body ionization since even a contact binary is not bound enough
compared to the typical velocity of a field star (Sec.\ref{ss:BinColl}).
The high velocity affects also 3-body interactions of a binary with
the MBH, since if the binary is tidally separated, one of the two
stars is ejected as a hyper velocity star (HVS) with velocity $\sim\sqrt{v\vs}$
\citep{hil88,yuq+03}, which can exceed the escape velocity from the
host galaxy (Sec. \ref{sss:TidalSeparation}).

\subsection{The special role of the Galactic MBH $\protect\SgrA$}

\label{ss:SgrA}

\begin{figure}
\noindent \begin{centering}
\includegraphics[width=0.75\columnwidth]{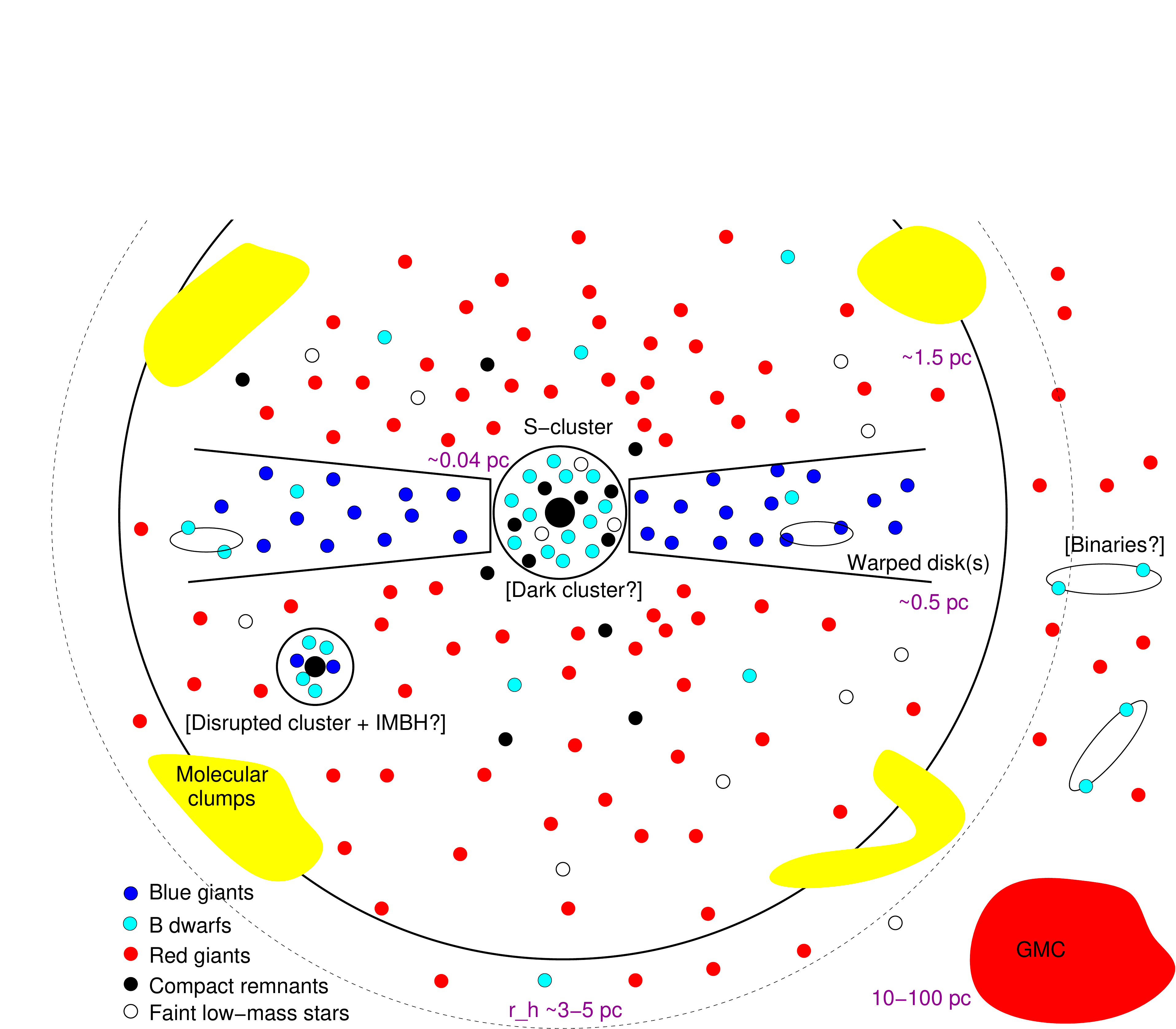}
\par\end{centering}
\caption{\label{f:MWGC}A not-to-scale schematic of the environment of the
Galactic MBH in the center of the Milky Way \citep{ale11}. Stars
are denoted by the small circles, color-coded by spectral type, binaries
are denoted by two stars on an ellipse, gas clouds are denoted by
irregular-shaped blobs. The existence of objects labeled by $[\ldots?]$
is unconfirmed or controversial. The MBH at the center (large black
circle) extends its direct dynamical influence up to $\protect\rh\sim3-5\,\mathrm{pc}$.
There is very little gas inside $r_{h}$, apart for a ring-like structure
of massive molecular clumps at $r\sim1.5\,\mathrm{pc}$. Further out
beyond $\protect\rh$ on the $10-100\,\mathrm{pc}$ scale there are
${\cal O}(10^{8})$ stars, multiple giant molecular clouds (GMCs)
and presumably also young binaries \citep{per+07}, which can be efficiently
scattered by the GMCs toward the MBH, and undergo tidal capture and
hyper velocity ejection (Sec. \ref{ss:3Body}). Inside $\protect\rh$,
the stellar population is a mix of old red giants in an approximately
isotropic distribution, and of young stars, mostly in two distinct
populations: ${\cal O}(100)$ massive OB stars in a warped, coherently
rotating stellar disk, extending between $\sim0.04\,\mathrm{pc}$
($1^{\protect\as}$) and $\sim0.5$ pc, and $\sim30$ MS B stars on
isotropic orbits inside $\sim0.04\,\mathrm{pc}$ (the so-called ``S-star
cluster''). It is still unclear what is the distribution of the long-lived,
low-mass faint MS stars, which are currently below the detection threshold,
and of the dark stellar remnants, in particular the ${\cal O}(10\,\protect\Mo)$
stellar BHs (the hypothetical ``dark cusp''). The presence of an
IMBH companion to $\protect\SgrA$, if light and distant enough, cannot
yet be excluded \citep{gil+09}. }
\end{figure}

$\SgrA$, the $\Mbh\simeq4\times10^{6}\,\Mo$ MBH at the center of
the Milky Way, $R_{0}\simeq8\,\mathrm{kpc}$ from the Solar system
\citep{gil+09,boe+16}, is the nearest MBH to us. Closer by a factor
$\sim100$ than the MBH in Andromeda, and by a factor $\sim2000$
than MBHs in the nearest galaxy cluster Virgo, $\SgrA$ is by far
the most accessible to high resolution observations, in spite of the
very strong dust extinction on the line of sight through the Galactic
disk \citep{sch+07}. Major instrumental and observational advances
in the IR and X-ray over the last few decades have made it possible
to observe faint accretion emission from the MBH and accurately track
individual stars and gas clumps by imaging and spectroscopy as they
orbit around it \citep[See review by][]{gen+10}. The wealth of detail,
the precision and the depth of observations of $\SgrA$ (\textbf{Figure}
\ref{f:MWGC}) are unlikely to be matched for other MBHs any time
soon. The Copernican principle suggests that galactic nuclei in general
are as complex as the Galactic Center. Therefore, the MBH at the center
of the Milky Way plays crucial roles as a source of information on
the environment of MBHs, as a driver of efforts to understand galactic
nuclei, and as a testbed for developing theories and models.

The central 1 pc of the Galactic Center contains a mixture of low-mass
red old stars (only the red giants are currently detectable) and massive
blue stars, the products of recent star formation \citep{bar+10}.
Many of the blue stars orbit in a coherent disk-like configuration
\citep{lev+03}, and appear to have been formed $\sim6$ Myr ago from
a fragmenting, self-gravitating circumnuclear gas disk, with an unusually
flat top-heavy initial mass function, quite unlike that observed in
star forming regions in the galactic bulk \citep{nay+05a,bar+10,luj+13}.
The observed blue OB giants will eventually explode as SNe, and populate
the Galactic Center with compact remnants (SNe remnants are still
seen in the region, \citealp[e.g.][]{mae+02}). There is further evidence
that episodic top-heavy star-formation has been going on in the inner
nucleus over the last 12 Gyr at a roughly constant rate \citep{man+07}.
This implies that the local stellar population that interacts with
the MBH may be quite different than that of the field.

The Galactic MBH is near the lower limit of MBH masses detected to
date \citep{kor+13}, in keeping with the $\Mbh/\sigma$ relation
and the relatively low-mass bulge of the Milky Way. The paucity of
$\SgrA$-like MBHs may reflect the strong selection bias against low-mass
MBHs. By its relatively low mass, $\SgrA$ is expected to be surrounded
by a dense cusp of dynamically relaxed old stars (Sec. \ref{ss:NR}).
This, however is not observed in the distribution of the red giants
($K\lesssim16^{\text{m}}$) \citep{do+09,buc+09,bar+10}, which are
currently the only directly detectable tracers of the faint old population.
Instead of rising, the red giant distribution is observed to flatten,
or even dip toward the center. Whether this implies that the faint
old population and its associated compact remnants is indeed missing,
for example, ejected by a fairly recent major merger \citep{mer10},
or whether this is merely the result of a selective destruction of
red giant envelopes \citep[e.g.][]{ama+14} is still unclear. However,
the low mass of $\SgrA$, the absence of other evidence for a past
major merger, and the fast relaxation expected by the presence of
stellar BHs in the system (Sec. \ref{ss:NR}; \citealt{pre+10}),
argue against the likelihood of a major merger in the recent past
of the Milky Way, or the possibility that an earlier one left the
system out of dynamical equilibrium up to the present day \citep{ale11}.
It should noted that a high density ``dark cusp'' of stellar remnants
(primarily stellar BHs) can develop on timescales that are much shorter
than the overall relaxation time \citep{pre+10,ant+14}. 

The question of the stellar density and the dynamical state around
$\SgrA$ is important because the rate of strong interactions with
the MBH scales with the stellar density close to it. In particular,
the Milky Way serves, by a coincidence of technology\footnote{The longest laser interferometric baseline that can be reliably maintained
in space at present is ${\cal O}(10^{6}\,\mathrm{km})$, which has
maximal sensitivity to GWs in the $1-10\,\mathrm{mHz}$ range. This
corresponds to GWs emitted from near the horizon of a $\mathrm{{\cal O}}(10^{6}\,\Mo)$
MBH \citep{ama+07}. }, as the archetypal nucleus for cosmic sources of low frequency GWs
from stellar BHs inspiraling into a MBH (extreme mass-ratio inspiral
events, EMRIs), which are targets for planned space-borne GW detectors\footnote{The central role of the Milky Way in the planning of such experiments
is reflected by the fact that the target galactic nuclei are sometimes
denoted Milky Way equivalent galaxies (MWEGs).}. The numbers and dynamics of stars and compact objects in the inner
$\mathrm{few\times0.01}$ pc of the $\SgrA$ therefore have, by extrapolation,
direct bearings on the predicted cosmic low-frequency GW event rates
(Sec. \ref{ss:EMRIs}).

A recent re-analysis of deep number counts ($K\sim18^{\text{m}}$,
dominated by old sub-giants, \citealp[Fig. 16]{sch+07}) and of the
diffuse light distribution, shows a cusp that rises from $\sim3\,\mathrm{pc}$
to the inner $\mathrm{few}\times0.01$ pc, with $\rho_{\star}\sim10^{7}(r/0.1\,\mathrm{pc})^{-1.20}\,\Mo\,\mathrm{pc^{-3}}$
\citep{sch+16}. If confirmed, this would imply that the missing cusp
problem is actually a problem of the missing brighter red giants in
the inner $<0.1$~pc, and that these do not trace the old faint population.

Finally, the Galactic Center offers a unique opportunity to detect
phenomena whose observational signature is presently too subtle to
be observed in more distant nuclei. These include very close star-star
encounters that lead to a stochastic spin-up and mixing of long-lived
stars near MBHs \citep{ale+01a}; ``Near-miss'' tidal interactions
between stars and the MBH, which result either in strong tidal scattering
that perturbs the internal structure and affects the evolution of
a substantial fraction of stars inside $r_{h}$ \citep{ale+01b,man+13},
or to orbital decay as ultra-luminous tidally powered ``squeezars''
\citep{ale+03a}, or tidal stripping of tenuous envelopes of extended
stars (a possible interpretation of the tidally sheared G2 cloud,
\citealt{gil+12,phi+13,gui+14}). A prime objective of observations
of $\SgrA$ and its environment is to detect subtle post-Newtonian
effects and test strong gravity (Sec. \ref{ss:GRorbits}).

\subsection{Scope of this review}

This review focuses on galactic nuclei that harbor a single MBH in
the lower mass range ($\Mbh<10^{8}\,\Mo$), and on the role stellar
dynamics play in enabling and regulating the unusual stellar phenomena
that occur near the MBH. 

There are theoretical and practical reasons to focus specifically
on this sub-class of galactic nuclei, beyond the fact that lower mass
galactic nuclei, which are associated with lower mass MBHs, are the
most numerous \citep{liy+11}. Such systems, if they evolve in isolation,
can reach dynamical quasi-equilibrium by stellar 2-body relaxation
\citep{bar+13,pre+10}. In that case, the steady state configuration
is independent of the unknown initial conditions, and so it is possible,
ideally, to deduce it from first principles, and to apply results
generally to all relaxed nuclei. Since the quasi-steady state is long-lived,
relaxed galactic nuclei are statistically the most likely to be observed
(assuming the relaxation time is substantially shorter than the age
of the universe). Note however that there is no guarantee that dynamically
relaxed galactic nuclei form a considerable fraction of the population;
the assumption of isolated evolution is a matter of boundary conditions,
which like initial conditions, are generally hard to determine reliably
or justify robustly in astrophysical contexts. Another unrelated physical
property that characterizes MBHs with mass below $10^{8}\,\Mo$ and
makes them more interesting is that they tidally disrupt MS stars
outside of the event horizon (Sec. \ref{p:TidalField}), which allows
the eruption of an observable tidal flare. 

There are also practical reasons to focus of nuclei with $\Mbh<10^{8}\,\Mo$.
The overlap of this mass range with the Galactic MBH $\SgrA$ means
that extrapolations of its properties are more likely to be justified,
and the technological coincidence of this mass range with the sensitivity
of planned space-borne interferometric GW detectors (Sec. \ref{ss:SgrA})
makes it of particular relevance for GW physics. 

Other recent reviews that complement and expand the material covered
here are \citet{ale05}: A theoretical review on the stellar processes
in the Galactic Center; \citet{gen+10}: An observational review on
the Galactic Center; \citet{gra16}: The MBH\textendash host galaxy
connection; \citet{mil+04b}: Intermediate mass BHs (IMBHs); \citet{ama+07}:
Detection of GW from EMRIs by space interferometers; \citet{mer+05}:
Binary MBHs; \citet{vol10}: Formation of MBHs. A comprehensive text-book
level coverage of stellar dynamics on galactic and nuclear scales
can be found in \citet{bin+08} and \citet{mer13}.

Acronyms commonly used in this review are listed for convenience in
table \ref{t:acronym}.
\begin{center}
{\footnotesize{}}
\begin{table}
{\footnotesize{}\tabcolsep7.5pt}{\footnotesize \par}

{\footnotesize{}\caption{\label{t:acronym}Acronyms commonly used in this review (by alphabetical
order)}
}{\footnotesize \par}

{\footnotesize{}}%
\begin{tabular}{>{\raggedright}p{0.08\columnwidth}>{\raggedright}p{0.35\columnwidth}|>{\raggedright}p{0.08\columnwidth}>{\raggedright}p{0.34\columnwidth}}
 & \multicolumn{1}{>{\raggedright}p{0.35\columnwidth}}{} &  & \tabularnewline
\hline 
{\footnotesize{}Acronym} & {\footnotesize{}Meaning} & {\footnotesize{}Acronym} & {\footnotesize{}Meaning}\tabularnewline
\hline 
{\footnotesize{}AGN} & {\footnotesize{}Active galactic nucleus / nuclei} & {\footnotesize{}ISCO} & {\footnotesize{}Innermost stable circular orbit}\tabularnewline
{\footnotesize{}AI} & {\footnotesize{}Adiabatic invariance} & {\footnotesize{}MBH} & {\footnotesize{}Massive black hole}\tabularnewline
{\footnotesize{}BH} & {\footnotesize{}Black hole} & {\footnotesize{}MS} & {\footnotesize{}Main sequence}\tabularnewline
{\footnotesize{}DF} & {\footnotesize{}Distribution function} & {\footnotesize{}NR} & {\footnotesize{}Non-resonant (2-body) relaxation}\tabularnewline
{\footnotesize{}EMRI} & {\footnotesize{}Extreme mass ratio inspiral} & {\footnotesize{}NS} & {\footnotesize{}Neutron star}\tabularnewline
{\footnotesize{}FP} & {\footnotesize{}Fokker-Planck (equation)} & {\footnotesize{}(S/V)RR} & {\footnotesize{}(Scalar / Vector) resonant relaxation}\tabularnewline
{\footnotesize{}GMC} & {\footnotesize{}Giant molecular cloud} & {\footnotesize{}SB} & {\footnotesize{}Schwarzschild barrier}\tabularnewline
{\footnotesize{}GR} & {\footnotesize{}General relativity} & {\footnotesize{}sma} & {\footnotesize{}Semi-major axis}\tabularnewline
{\footnotesize{}GW } & {\footnotesize{}Gravitational wave(s)} & {\footnotesize{}SNe} & {\footnotesize{}Supernovae}\tabularnewline
{\footnotesize{}HVS} & {\footnotesize{}Hyper velocity star} & {\footnotesize{}TDE} & {\footnotesize{}Tidal disruption event}\tabularnewline
{\footnotesize{}IMBH} & {\footnotesize{}Intermediate mass black hole} & {\footnotesize{}WD} & {\footnotesize{}White dwarf}\tabularnewline
\hline 
\end{tabular}{\footnotesize \par}
\end{table}
\par\end{center}{\footnotesize \par}

\section{Dynamical relaxation near a MBH}

\label{s:dynamics}

Close enough to the MBH, inside the radius of influence, $r_{h}$,
where the total stellar mass is small, $\Ns\Ms/\Mbh\ll1$, but far
enough that the system is Newtonian, $r/r_{g}\gg1$, the mass of the
MBH dominates the potential and the dynamics are approximately Keplerian
over timescales much longer than the Keplerian orbital timescale $P=2\pi\sqrt{a^{3}/G\Mbh}$\footnote{The typical radius $r$, the Keplerian semi-major axis $a$, and the
circular radius corresponding to orbital energy $E$, are used here
interchangeably in approximate derivations and relations.}. The discussion here focuses on how such a system ``forgets'' its
initial conditions, and approaches quasi-steady state. The true steady
state of a self-gravitating system is that of maximal inhomogeneity:
a small dense central system (a binary) and an unbound halo at infinity
\citep[e.g.][]{bin+08}. The evolutionary time to achieve this steady
state is usually much longer than the Hubble time $t_{H}$. It is
more meaningful to consider the quasi-steady state of the system over
$t_{H}$, which for brevity will be denoted here steady-state. 

The primary parameters that determine the steady state are the ratio
of MBH mass to the mean stellar mass, $Q=\Mbh/\left\langle \Ms\right\rangle $
and $N_{h}$, the total number of stars enclosed within $r_{h}$.
The secondary parameters are the rms of the stellar mass function,
$\left\langle \Ms^{2}\right\rangle ^{1/2}$ and the stellar density
distribution inside $r_{h}$, which is approximated here as a power-law
with logarithmic slope $\alpha$, so that $\Ns(<r)=N_{h}(r/r_{h})^{3-\alpha}$
and $\ns(r)\propto r^{-\alpha}$. 

Two-body relaxation is inherent to motion in a potential generated
by discrete objects. In the impulsive limit, the force a perturbing
star exerts on a test star at nearest approach, over a small fraction
of the test star's orbit, deviates substantially from the superposed
force of the entire system, which governs motion on timescales shorter
than the relaxation time. Over time, the cumulative effect of such
uncorrelated 2-body gravitational interactions randomizes both the
orbital energy and angular momentum. This slow but unavoidable non-coherent
collisional relaxation process\footnote{Partial randomization occurs also in collisionless systems by phase-mixing
(including violent relaxation and Landau Damping) and chaotic mixing.
However, it is yet not well understood what are the general properties
of the end states of these processes \citep{moh+10}.} can be described as diffusion in phase space in terms of the Fokker-Planck
(FP) equation (Sec. \ref{ss:NR}). In the opposite limit of very soft
encounters, where the nearest approach of the perturbing star is larger
than the orbit of the test star, it is no longer valid to describe
the interaction as occurring instantaneously and locally between two
stars. Rather, the long-term temporal correlations of the perturbing
background in the near-symmetric potential near a MBH drive a different
form of randomization, coherent (resonant) relaxation (Sec. \ref{ss:RR}).
These two limits of relaxation can be incorporated in a unified framework
(Sec. \ref{ss:GenRelax}, \textbf{Table} \ref{t:rlx}). 

\subsection{Non-coherent Relaxation}

\label{ss:NR}

\subsubsection{2-body relaxation}

\label{sss:2Body}

A test star orbiting with semi-major axis (sma) $a$ around a central
mass $\Mbh$ surrounded by stars of mass $\Ms$ and space density
$\ns$, is subject to a net residual specific force $\sqrt{\left\langle F^{2}\right\rangle }\sim\sqrt{\mathrm{d}\Ns}G\Ms/b^{2}$
from the $\mathrm{d}\Ns\sim\ns b^{2}\mathrm{d}b$ stars in a thin
shell of radius $b\ll a$ around it, due to the Poisson fluctuations
about the mean density. The residual force persists until the stars
generating it move substantially, after a short coherence time $T_{c}^{NR}\sim b/\sigma\ll a/\sigma\sim P(a)$,
where $\sigma^{2}\sim G\Mbh/a$ is the velocity dispersion. Because
$T_{c}^{NR}\ll P$, the encounter is impulsive (a collision). The
short duration of the interaction and the fact that successive collisions
involve different perturbers, justify the assumption that the perturbations
are a Markovian (memory-less) process, and this allows their cumulative
effect to be described as diffusion . The diffusion timescale for
both energy and angular momentum is $T_{E}=E^{2}/D_{EE}\sim J_{c}^{2}/D_{JJ}$,
where $J_{c}=\sqrt{G\Mbh a}$ is the circular angular momentum, and
$D_{EE}$ and $D_{JJ}$ are the energy and angular momentum diffusion
coefficients. 

It is also useful to define relaxation in terms of quantities in position
and velocity space. The impulse to the test star by a single collision
is $\delta v\sim\sqrt{\left\langle F^{2}\right\rangle }T_{c}^{NR}$.
These accumulate non-coherently over times $t>T_{c}^{NR}$ to $\left\langle \Delta v^{2}\right\rangle \sim(G^{2}\Ms^{2}\ns/\sigma)\mathrm{d}b/b$.
Integration over all shells from $b_{\min}$ to $b_{\max}$ assuming
constant $\ns$ yields the non-coherent (2-body) relaxation time,
also known as the Chandrasekhar time, 
\begin{equation}
T_{NR}(r)=\frac{\sigma^{2}}{\left\langle \Delta v^{2}\right\rangle }\sim\frac{\sigma^{3}}{G^{2}\Ms^{2}\ns\log\Lambda}\sim\frac{Q^{2}P(r)}{\Ns(r)\log Q}\,,\label{e:TNR}
\end{equation}
where $\Lambda=b_{\max}/b_{\min}$ is the Coulomb factor, and the
last approximate equality holds when the system is dominated by a
central MBH, in which case $\Lambda\sim Q$ \citep{bar+13}. The Coulomb
logarithm lies between $14\lesssim\log Q\lesssim23$ over the known
range of MBH masses. $T_{NR}\propto r^{\alpha-3/2}$ is typically
a weak function of radius when a stellar cusp surrounds the MBH (Sec.
\ref{sss:SScusp}). 

The time for the stellar system around a MBH to recover from a perturbation
and return to steady state can be expressed in terms of quantities
at the radius of influence as \citep{bar+13} 
\begin{equation}
T_{\mathrm{ss}}\simeq10T_{E}(r_{h})\simeq\frac{5}{32}\frac{Q^{2}P_{h}}{N_{h}\log Q}\,.\label{e:Tss}
\end{equation}
 Although the diffusion timescale $T_{E}(E,J)$, the Chandrasekhar
(NR) time $T_{NR}(r)$ and the relaxation time $T_{\mathrm{ss}}$
(at $r_{h}$) all express the tendency of the system to evolve toward
steady state, their numerical values near a MBH are quite different,
$T_{NR}(r_{h})\sim4T_{ss}\sim40T_{E}$ \citep{bar+13}. This distinction
matters because often $T_{\mathrm{ss}}$ is not clearly different
from the age of the system, and inferences about the dynamical state
then depend sensitively on the precise definitions and values of these
timescales.

\subsubsection{Anomalous diffusion}

\label{sss:AnomalousDiffusion}

Past attempts to measure the NR relaxation rate and the approach to
steady state in numeric experiments resulted in diverging estimates
of the dynamical state inside the radius of influence, especially
very close to the MBH (cf discrepancy between \citealt{bah+77} and
\citealt{mad+11}). The likely cause is that relaxation by gravitational
interactions is not a true diffusion process, but rather a marginal
case of anomalous diffusion \citep{bar+13}, which is very slow to
converge and therefore difficult and misleading to measure by standard
methods. 

The statistical evolution of the stellar orbital energy distribution
is expressed by the energy propagator function, which describes the
evolution of an initially mono-energetic distribution of stellar energies.
For true diffusion, the propagator converges quickly by the Central
Limit theorem to a Gaussian, whose width grows as $\sqrt{D_{EE}t}$.
In that limit the process is fully described by the two lowest order
diffusion coefficients (drift and scatter). The Markovian assumption
of independent scattering events is however not sufficient to ensure
the applicability of the Central Limit. In addition, the mean and
variance of the propagator must be finite. This loophole (diverging
moments) allows certain physical processes to display a very slowly
converging anomalous diffusion. In those cases the probability distribution
of the propagator is heavy-tailed; ``impossibly'' large changes
(when estimated naively by Gaussian distribution standards) actually
do occur surprisingly frequently, and all higher order diffusion coefficients
contribute to the evolution. Gravitational scattering happens to be
such an anomalous process with a formally diverging variance due to
the $r\to0$ divergence of the gravitational potential of a point
mass \citep{goo83}. This makes energy relaxation by gravitational
interactions a marginally anomalous diffusion process, with a heavy-tailed
distribution that initially grows as $\sim\sqrt{t\log(t)}$ \citep{bar+13}.
Since no physical process can truly diverge, the propagator must eventually
converge to a Gaussian. Careful analysis shows that this occurs well
before the MBH / nucleus system reaches steady state (Eq. \ref{e:Tss}).
However, it can affect the nature of gravitational perturbations on
short timescales. For example, it may be detected by future high angular
resolution observations of stellar orbits near $\SgrA$ \citep{bar+13}.
The relevance of anomalous diffusion for angular momentum evolution
remains to be studied. 

\subsubsection{Massive perturbers}

\label{sss:MPs}

The derivation of the 2-body relaxation timescale (Eq. \ref{e:TNR})
is easily generalized to the realistic case where the stellar population
has a spectrum of masses, $\Ms^{2}\ns\to\left\langle \Ms^{2}\right\rangle n_{\star}$.
Since $\left\langle \Ms^{2}\right\rangle \ge\left\langle \Ms\right\rangle ^{2}$,
a mass spectrum always implies faster relaxation than estimated by
naively substituting $\Ms\to\left\langle \Ms\right\rangle $. In particular,
relaxation can be extremely fast if the system contains a few massive
perturbers (for examples giant molecular clouds (GMCs), stellar clusters,
or IMBHs), with mass $M_{p}\gg\Ms$ \citep{spi+51}. In that case,
the quadratic dependence on the mass typically more than compensates
for the low space density of the massive perturbers, $n_{p}$, and
NR is accelerated by $T_{NR,\star}/T_{NR,p}\sim M_{p}^{2}n_{p}\left/\left\langle \Ms^{2}\right\rangle n_{\star}\right.\gg1$.
For example, GMCs in the Galactic center accelerate relaxation by
a factor of $10^{4}-10^{6}$ compared to relaxation by stars \citep{per+07}.
The effectiveness of a gravitational encounter with extended massive
perturbers (GMCs or clusters) is however suppressed logarithmically
by a factor $\log\Lambda_{\mathrm{extended}}/\log\Lambda_{\mathrm{point}}$,
since close (penetrating) encounters involve only a fraction of the
extended perturber's mass. For the relaxation of a point-like object
(a star, or a binary when treated as if concentrated at its center
of mass), this suppression is a small correction only. However, for
the relaxation-like process of binary evaporation (Sec. \ref{ss:BinColl}),
extended massive perturbers are inefficient when their size far exceeds
the binary sma. Note that the presence of massive perturbers is relevant
only when stellar NR alone is too slow to bring the system to full
randomization. An example of this distinction is the modest increase
due to massive perturbers in the stellar TDE rate (Sec. \ref{ss:TDEs}),
as opposed to the huge boost in the binary tidal separation rate \citep{per+07}
(Sec.\ref{sss:TidalSeparation}). 

\subsubsection{Steady state stellar cusp}

\label{sss:SScusp}

The steady state density distribution of a spherical Keplerian system
of single mass stars around a MBH is approximately a power-law cusp,
$\ns\propto r^{-7/4}$ (the distribution function (DF) is $f(E)\propto E^{1/4}$)
\citep{bah+77}. The Bahcall-Wolf solution averages over angular
momentum and describes the stellar current to the MBH, $I_{\bullet}$,
in energy only\footnote{The steady state Bahcall-Wolf solution can be derived qualitatively
by assuming that the orbital energy lost by stars that are scattered
into the MBH is carried outward by the stellar system on the NR timescale
\citep[e.g.][]{bin+08}.}. Stars fall in the MBH when they cross below some threshold $a_{\bullet}$.
This approximation captures quite well the solution in the intermediate
range $\rbh\ll r\ll r_{h}$, as verified by more detailed calculations
\citep[e.g.][]{bar+16}. The Bahcall-Wolf solution is similar to the
$I_{\bullet}=0$ (zero current) solution that holds in the limit of
a closed system, where the stellar current at $a_{h}$ is set to zero,
and the limit $a_{\bullet}/a_{h}\to0$ (no sink) is assumed. That
this solution also describes a realistic open system, appears at odds
with the very large natural scale for the stellar diffusion current
in the system, $I_{\star}\sim{\cal O}(\Ns/T_{E})$. However, this
is due to the fact that since $a_{\bullet}/a_{h}\ll1$, the drift
and scatter currents are near-zero\footnote{\label{fn:0current}Such a steady state solution, or one where $I_{\bullet}=0$
is realized by a cancellation between large drift and scatter currents,
require a delicate balance between the drift and scatter coefficients.
Slight deviations from the true form of the diffusion coefficients
(e.g. when they are not known exactly, and must be derived empirically
from $N$-body simulations, or from approximate physical arguments)
can lead to large errors in the steady state DF and/or unrealistically
large currents $I_{\bullet}\sim{\cal O}(I_{\star})$.}, and therefore $0<I_{\bullet}\ll I_{\star}$.

\subparagraph{Dynamical friction}

Two-body NR scattering (described by the 2nd order diffusion coefficient,
the scatter, which depends only on the mass spectrum of the background
stars since all test masses accelerate similarly) is balanced by dynamical
friction (described by the 1st order diffusion coefficient, the drift,
which is proportional to the test mass since it determines how massive
is the wake that forms behind the test star and drags it). The deceleration
due to the drift of a massive test mass $M\gg\left\langle \Ms\right\rangle $
moving slowly relative to the background, scales with $v$ \citep[e.g.][]{bin+08},
and so dynamical friction acts like viscous dissipation. As the test
mass slows down, it sinks further in the potential until it either
``runs out'' of stars (the stellar mass enclosed in its orbit is
smaller than $M$) \citep[e.g.][]{mer13}, or it reaches the tidal
disruption radius or event horizon and is destroyed.

\subparagraph{Mass segregation}

Realistic stellar populations have a broad mass spectrum. Dynamical
friction in the steep MBH potential frustrates the trend toward equipartition,
since when the massive stars slow down, they sink inward. Conversely,
light masses speed up and migrate to wider orbits. Mass segregation
modifies the single mass power-law Bahcall-Wolf solution. Every mass
bin $\Ms\mathrm{d}\Ms$ is characterized by its own power-law density
profile $n_{\Ms}\propto r^{-\alpha_{\Ms}}$ where $\alpha_{\Ms}$
is larger the more massive the star, with $3/2\lesssim\alpha_{\Ms}\lesssim5/2$
\citep{bah+77,ale+09,kes+09} The quasi steady state of a mass segregated
cusp can still be approximated by a Maxwell-Boltzmann mass-dependent
velocity distribution \citep{ale+01a}, but the spread in velocities
is much reduced compared to equipartition, $\sigma_{\Ms}^{2}(r)=G\Mbh/(1+\alpha_{\Ms})r$
\citep{ale99}.

\subparagraph{Weak vs strong mass segregation}

\begin{figure}
\centering{}\includegraphics[width=1\textwidth]{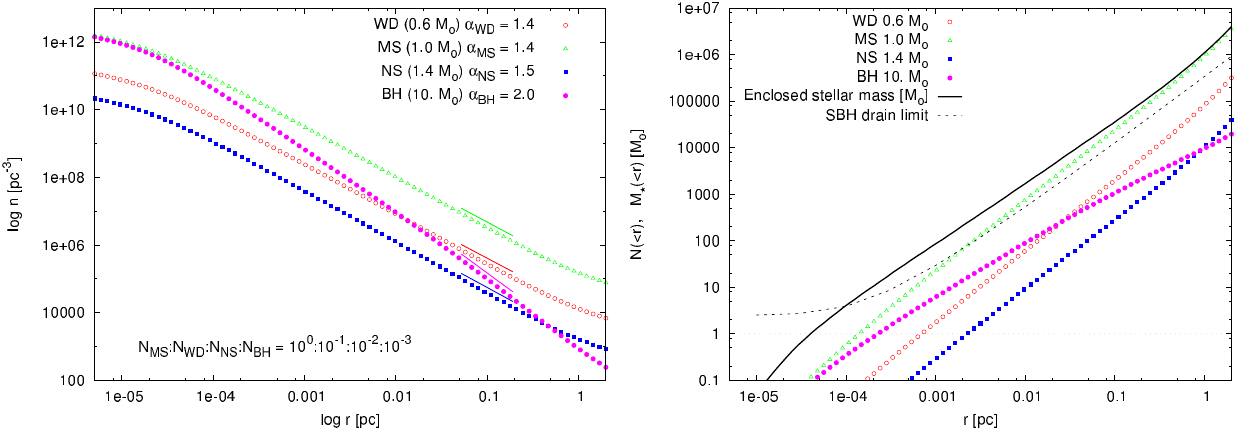}\caption{\label{f:MSeg}The effects of strong mass segregation on the density
distribution of a simplified population with MS stars and remnants
(WDs, NSs and stellar BHs) around a Milky Way-like nucleus with $\protect\Mbh=4\times10^{6}\,\protect\Mo$
\citep{ale+09}. \textbf{Left}: The number density and approximate
power-law indices for the distributions of stars and remnants. \textbf{Right}:
The enclosed mass as function of radius, and the drain limit upper
bound for stellar BHs (Eq. \ref{e:drainlim}).}
\end{figure}

The mass segregation steady state solution has two branches, depending
on the stellar mass function. In the \citet{bah+77} weak segregation
solution the approximate power-law density distribution has a mass-dependent
logarithmic slope, $\alpha(\Ms)=3/2+(\Ms/\max\Ms)/4$. Thus the heaviest
stars assume the $\alpha_{H}=7/4$ power-law of a single mass population,
while the lightest stars have a somewhat less centrally concentrated
distribution with $\alpha_{L}=3/2$ ($\alpha_{H}-\alpha_{L}=1/4$).
The full range of steady state mass segregation solutions depends
on the mass function of the stellar population \citep{ale+09}. Long-lived
stellar populations, whether old star-bursts or continually star-forming
ones, are well approximated by a two-mass population: the $M_{L}\sim{\cal O}(1\,\Mo)$
light stars with a number density $N_{L}$ in the unsegregated unbound
population, which includes low mass MS stars, WDs and neutron stars
(NSs), and the $M_{H}\sim{\cal O}(10\,\Mo)$ heavy remnants, with
number density $N_{H}$, which consists of stellar BHs. The nature
of the mass segregation solution depends on the value of the the relaxational
parameter $\Delta=4N_{H}M_{H}^{2}\left/\left[N_{L}M_{L}^{2}(3+M_{H}/M_{L})\right]\right.$,
which measures the relative strength of heavy/heavy scattering to
heavy/light dynamical friction \citep{ale+09}. The weak segregation
solution applies when $\Delta\gg1$. In that case the heavy stars
are the dominant component of the population, interactions with the
light stars are negligible, and therefore the heavy stars behave as
a single-mass population. The strong segregation solution applies
when $\Delta\ll1$. In that case the heavy stars are a trace component
of the population, and therefore they interact mainly with the light
stars, and are driven efficiently to the center by dynamical friction\footnote{This is different, and in the opposite sense from the  \citet{spi69}
instability in a cluster without a MBH, where a dominant stellar BH
population segregates and forms a dense central subsystem. }. The strong segregation solution leads to steeper slopes and a larger
difference between the light and heavy masses, $2\lesssim\alpha_{H}\lesssim11/4$
and $3/2<\alpha_{L}<7/4$ ($\alpha_{H}-\alpha_{L}\simeq1$). The general
mass segregation solution for an arbitrary mass function can have
an even steeper density profile slope, $\alpha(\max\Ms)=5/2$ \citep{kes+09}.
These analytic results were confirmed by $N$-body simulations \citep{pre+10}.
Long-lived stellar populations have $\Delta<0.1$, and therefore many
relaxed galactic nuclei are expected to be strongly segregated, and
have steep high density cusps of stellar BHs. This is true in particular
for the Galactic center \citep{mor93}.\textbf{ Figure} \ref{f:MSeg}
shows the strong mass segregation that is predicted for the stellar
BHs in a Milky Way-like galactic nucleus. Strong mass segregation
of stellar BHs increases the EMRI rate by a factor $\times10$ by
concentrating more BHs below the critical sma for inspiral (Sec. \ref{sss:PvsI}).

\subparagraph{The drain limit}

A useful upper limit on the number of stars that can exist in quasi-steady
state inside a small volume around a MBH can be obtained by requiring
that the mean time for 2-body interactions to scatter a star into
the MBH be longer than the age of the system. The drain limit \citep{ale+04}
is given by the condition $\Ns^{-1}\mathrm{d}\Ns/\mathrm{d}t=\left[\log(\sqrt{2r/r_{g}})T_{NR}(r)\right]^{-1}=t_{H}^{-1}$
(Sec. \ref{ss:NewtonLC}). The solution, $\max\Ns(r$), is the maximal
number of stars that can be initially packed inside $r$ with a survival
probability of $1/2$ over $t_{H}$ (assumed to be the age of the
system),
\begin{equation}
\max\Ns(<r)\sim\frac{2}{3}\frac{\log\sqrt{2r/r_{g}}}{\log Q}\frac{\Mbh^{2}}{\left\langle \Ms^{2}\right\rangle }\frac{P(r)}{t_{H}}\,.\label{e:drainlim}
\end{equation}
The dependence on $\left\langle \Ms^{2}\right\rangle ^{-2}$ reflects
the acceleration of the relaxation process in the presence of a mass
spectrum (Sec. \ref{sss:MPs}). Comparison with detailed calculations
\citep[e.g.][]{dee+07} shows that the drain limit is a somewhat conservative
upper bound (by up to a factor of a few at $r\ll r_{h}$, cf \textbf{Figure}
\ref{f:MSeg}). 

\subsubsection{Dynamically relaxed galactic nuclei}

\label{sss:RelaxedNuclei}

The empirical $\Mbh/\sigma$ relation (Sec. \ref{ss:host}), $\Mbh\propto\sigma^{\beta}$,
implies a scaling for the 2-body relaxation time of galactic nuclei.
Since the MBH's radius of influence scales as $r_{h}\sim G\Mbh/\sigma^{2}\propto\Mbh^{1-2/\beta}$
and the number of stars inside $r_{h}$ scales as $N_{h}\propto Q$,
the time to reach steady state (Eq. \ref{e:Tss}) at the radius of
influence scales as $T_{ss}\propto Q^{2}P_{h}/N_{h}\propto\Mbh^{2-(3/\beta)}$
(omitting the logarithmic Coulomb factor) \citep{ale11}. The relaxation
time thus increases with the MBH mass for $\beta>1.5$. The $\Mbh/\sigma$
relation also implies that the mean stellar density inside $r_{h}$
scales as $\bar{n}_{\star}\sim3Q/4\pi r_{h}^{3}\propto\Mbh^{(6/\beta)-2}$,
which falls with the MBH mass for $\beta>3$. The empirically determined
range $\beta\sim4-5$ then means that isolated galactic nuclei with
lower mass MBHs have denser, more relaxed nuclei. A detailed study
of the relaxation process, which fixes the exact numeric prefactors
that enter $T_{ss}$ \citep{bar+13} shows that galactic nuclei with
$\Mbh<\mathrm{few\times}10^{7}$, which evolve passively in isolation,
should be dynamically relaxed by $t_{H}$. This conclusion is further
reinforced when the accelerating effect of massive stars is taken
into account \citep{pre+10}.

\subsection{Coherent (resonant) relaxation}

\label{ss:RR}

Resonant Relaxation (RR) is a rapid angular momentum relaxation mechanism
that operates in symmetric potentials, which restrict orbital evolution
\citep{rau+96,hop+06a}. In particular, an MBH enforces a high degree
of symmetry on the gravitational potential at $r\ll r_{h}$. In the
Keplerian limit, where both star-star interactions and GR are negligible,
stars move on fixed ellipses around the MBH. Assume that the orbital
parameters of a test star are statistically similar to those of the
background (this is later relaxed in Sec. \ref{ss:relLC}). Even when
the stellar distribution is spherical and isotropic on average, Poisson
fluctuations due to the finite number of stars generate a non-zero
residual force field that acts on the test star, with rms $F_{N}\sim\sqrt{\Ns(a)}G\Ms/a^{2}$.
The restricted orbital evolution on timescales much longer than the
radial orbital period $P$ allows the phase-averaging of the residual
force over $P$ (the mean anomaly), of both the test star and the
background. This implies that the phase-averaged force conserves the
orbital energy of the test star\footnote{\label{fn:avH}This is formally expressed by double averaging the
Hamiltonian that describes the MBH/star and the star/star interactions
in terms of action-angle canonical coordinates, over the mean anomalies
of the test star and background stars. The conjugate action of the
test star $I=\sqrt{G\Mbh a}$ is then conserved, and therefore so
is its orbital energy \citep{bar+14}.}. In the case of a nearly Keplerian system, the phase-averaged orbits
can be conceptualized as fixed mass wires, whose total mass is that
of the star, and whose linear mass density is inversely proportional
to the local orbital velocity.

The residual rms torque on the test star\footnote{\label{fn:RRrcorr}The effective extent of an $\ns\propto r^{-\alpha}$
background that exerts a coherent torque on a star with sma $a$ is
$\lesssim2^{2/(3-\alpha)}a$ \citep{bre+09}, as demonstrated numerically
by \citet{gur+07}. For typical values of $\alpha$, the spatial coherence
of the RR torques is $\Delta\log a\sim1$.}, $\tau_{N}\sim aF_{N}\sim\sqrt{\Ns(a)}G\Ms/a$, persists as long
as the background orbits remain fixed, over a coherence time $T_{c}$.
The coherence time is limited by the fastest process that breaks the
symmetry and randomizes the background. The relevant processes for
a symmetric stellar cusp around a non-spinning MBH are the retrograde
precession of the argument of periastron due to the enclosed stellar
mass inside the orbit (``mass precession''), which drops to zero
for $j\to0$ (radial) orbits and grows larger as $a\to r_{h}$, in-plane
prograde GR precession of the argument of periastron (``Schwarzschild
precession''), which is larger the smaller the orbital periapse,
that is, larger for $j\to0$ and $a\to0$ orbits, and ultimately,
the residual torques themselves randomize the background orbits (``self-quenching''),
since any star in the system can play the role of a test star. The
mass precession coherence time is $T_{c}^{M}(a)\sim QP(a)/\Ns(a)$;
the isotropically averaged GR precession time is $T_{c}^{GR,\mathrm{iso}}(a)=(a/12r_{g})P(a)$;
the self-quenching coherence time is $T_{c}^{SQ}(a)\sim QP(a)/\sqrt{\Ns(a)}$
\citep[e.g.][]{hop+06a}. The exact numeric prefactors relating the
system parameters to the resonant torques and coherence times are
difficult to derive analytically, but they can be calibrated by simulations
\citep{rau+96,eil+09,koc+15,bar+16}.

On times shorter than $T_{c}$, the change in the angular momentum
of the test star grows linearly with time up to $(\Delta\boldsymbol{J})_{c}=\boldsymbol{\tau}_{N}T_{c}$.
This maximal coherent change then becomes the mean free path for a
random walk in angular momentum phase-space on times longer than $T_{c}$,
$\left|\Delta\boldsymbol{J}(t)\right|=\left|(\Delta\boldsymbol{J})_{c}\right|\sqrt{t/T_{c}}$.
It is convenient to normalize the evolution in angular momentum to
$J_{c}(a)$, and define the relaxation timescale as the time for the
relative change to reach order unity, $T_{RR}=J_{c}^{2}/(\tau_{N}^{2}T_{c})$,
so that $\Delta j(t)=\sqrt{t/T_{RR}}$. In terms of the system parameters
(cf Eq. \ref{e:TNR}),
\begin{equation}
T_{RR}(a)=\frac{J_{c}^{2}(a)}{\tau_{N}^{2}(a)T_{c}(a)}\sim\frac{Q^{2}}{\Ns(a)}\frac{P^{2}(a)}{T_{c}(a)}\sim\left[\frac{P(a)}{T_{c}(a)}\log Q\right]T_{NR}(a)\,.\label{e:TRR2TNR}
\end{equation}
Unlike NR, where 2-body relaxation is boosted by diverging local point-point
interactions, as expressed by the Coulomb factor, RR proceeds by the
interaction of extended objects (the mass wires), whose mutual torques
do not diverge, but are rather boosted by the long coherence time.
RR is significant for dynamics near a MBH because the coherence time
there can be many orders of magnitude longer than the orbital time,
whereas $\log Q\sim{\cal O}(10)$, and therefore $T_{RR}\ll T_{NR}$
(Eq. \ref{e:TRR2TNR}). Since $j\to0$ orbits lead to strong interactions
with the MBH, rapid angular momentum relaxation by RR can potentially
dominate the dynamics leading to such interactions.

The residual torque due to the superposed forces of many wires changes
both the direction and magnitude of the test star's angular momentum.
Such a resonant relaxation process is denoted (confusingly) ``scalar
RR'' (SRR), to emphasize that it changes $j$, and in particular
can drive it to $j\to0$. The coherence time of SRR is set by the
combined in-plane retrograde mass precession and prograde GR precession,
$T_{c}^{\mathrm{prec}}\sim\left|1/T_{c}^{M}-1/T_{c}^{GR}\right|^{-1}$.
In the limit where $T_{c}^{\mathrm{prec}}=T_{c}^{M}$, the RR relaxation
timescale is $T_{RR}^{M}(a)\sim QP(a)$ (Eq. \ref{e:TRR2TNR}).

SRR is to be contrasted with the residual torques that arise on timescales
much longer than $T_{c}^{\mathrm{prec}}$, when the orbital rosettes
can be averaged over the precession period and conceptualized as concentric
mass annuli. On those longer timescales, the residual torque on a
test annulus due to the Poisson fluctuations in the orbital orientations
of the finite number of background annuli has, by symmetry, only a
transverse component. This changes the test annulus orientation, but
not the rosette's eccentricity ($j$). This restricted form of RR
is called ``vector RR'' (VRR), and its coherence time is set by
self-quenching. By its definition (Eq. \ref{e:TRR2TNR}), $T_{RR}^{SQ}\sim T_{c}^{SQ}\sim QP(a)/\sqrt{\Ns(a)}$,
i.e., the self-quenching coherence time is similar to the time is
takes the orbital orientations to reach maximum randomization. 

\subsection{A general framework for describing relaxation}

\label{ss:GenRelax}
\begin{center}
{\footnotesize{}}
\begin{table}
{\footnotesize{}\tabcolsep7.5pt}{\footnotesize \par}

{\footnotesize{}\caption{\label{t:rlx}Hierarchy of relaxation processes}
}{\footnotesize \par}

{\footnotesize{}}%
\begin{tabular}{l|c|c|c}
\multicolumn{1}{l}{} & \multicolumn{1}{c}{} & \multicolumn{1}{c}{} & \tabularnewline
\hline 
{\footnotesize{}Process} & {\footnotesize{}NR} & {\footnotesize{}Scalar RR} & {\footnotesize{}Vector RR}\tabularnewline
\hline 
{\footnotesize{}Effective particles} & {\footnotesize{}Points} & {\footnotesize{}Ellipses} & {\footnotesize{}Annuli }\tabularnewline
{\footnotesize{}Averaged quantity} & {\footnotesize{}None} & {\footnotesize{}Mean anomaly } & {\footnotesize{}Arg. of periapse}\tabularnewline
{\footnotesize{}Conserved quantity} & {\footnotesize{}None} & {\footnotesize{}$E$} & {\footnotesize{}$E,J$}\tabularnewline
{\footnotesize{}Coherence time } & {\footnotesize{}$T_{c}^{NR}\!<\!P$} & {\footnotesize{}$P\!<\!T_{c}^{sRR}\!\sim T_{c}^{M}\!<\!T_{p}\,^{(a)}$} & {\footnotesize{}$T_{p}\,^{(a)}\!<\!T_{c}^{vRR}\!\sim\!T_{c}^{SQ}$}\tabularnewline
{\footnotesize{}Residual rms force } & {\footnotesize{}$\sqrt{\Ns}G\Ms\sqrt{Q}/a^{2}\,^{(b)}$} & {\footnotesize{}$\sqrt{\Ns}G\Ms/a^{2}$} & {\footnotesize{}$\sqrt{\Ns}G\Ms/a^{2}$}\tabularnewline
{\footnotesize{}Relaxation time $^{c}$} & {\footnotesize{}$Q^{2}P/\Ns\log Q\,^{(b)}$} & {\footnotesize{}$QP$} & {\footnotesize{}$QP/\sqrt{\Ns}$}\tabularnewline
\hline 
\multicolumn{4}{l}{{\footnotesize{}$^{{\rm a}}$~$T_{p}$ is the test star's precession
period.}}\tabularnewline
\multicolumn{4}{l}{{\footnotesize{}$^{{\rm b}}$~Integrated over all impact parameters.}}\tabularnewline
\multicolumn{4}{l}{{\footnotesize{}$^{c}$~$T_{\mathrm{relax}}\sim J_{c}^{2}/\left\langle \tau_{N}^{2}\right\rangle T_{c}$.}}\tabularnewline
\hline 
\end{tabular}{\footnotesize \par}
\end{table}
\par\end{center}{\footnotesize \par}

Relaxation around a MBH can be described more generally in terms of
a hierarchy of coherence times, averaged and conserved quantities,
effective particles and symmetries (\textbf{Table} \ref{t:rlx}; \citealt{bar+16}).
The general picture is that for progressively slower coherence-breaking
mechanisms (the hierarchy: impulse$\to$in-plane precession$\to$self-quenching
torques) and correspondingly longer coherence times ($T_{c}^{NR}$$\to[$$T_{c}^{M}$,$T_{c}^{GR}]$$\to$$T_{c}^{SQ}$),
there are more periodic degrees of freedom in the problem that can
be averaged out (none$\to$mean anomaly$\to$angle of periapse). For
each of these emerges another conserved orbital quantity (none$\to$$E$$\to$$J$)
(footnote \ref{fn:avH}). With each successive averaging, the symmetry
of the effective particle grows (points$\to$wires$\to$annuli) and
therefore the number of the remaining ``torquable'' degrees of freedom
decreases, and the resonant torques on it decrease in magnitude. This
decrease in $\tau_{N}$ is however only by an order unity factor,
whereas the coherence time $T_{c}$ increases with symmetry as some
power of $\Ns$, which is a very large number. Therefore, the net
result is that the relaxation timescales $T_{RR}\propto1/(\tau_{N}^{2}T_{c})$,
decrease with higher symmetry ($T_{NR}>T_{sRR}>T_{vRR}$).

Non-coherent two-body relaxation fits naturally in this framework
as the limiting case of minimal symmetry, shortest coherence time
and longest relaxation timescale. NR is treated in the impulsive limit,
where the interaction is effectively limited to the short flyby that
lasts much less than the orbital time. For that reason, no averaging
is possible and there are no conserved orbital quantities in the test
orbit\textemdash NR can change any of them. 

\begin{figure}
\noindent \begin{centering}
\includegraphics[width=0.65\columnwidth]{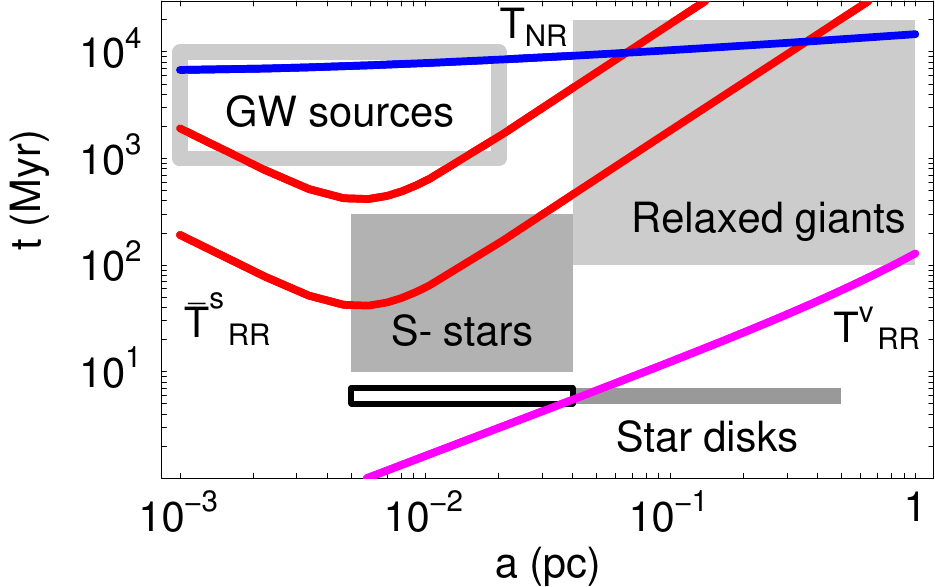}
\par\end{centering}
\caption{\label{f:GCrelax}The relaxation timescales in the Milky Way's nucleus
near the $4\times10^{6}\,\protect\Mo$ MBH, shown against the typical
ages of various stellar populations there, assuming a relaxed stellar
cusp \citep{hop+06a}. The non-resonant (2-body) relaxation time $T_{NR}$
(blue line) is nearly constant and only somewhat shorter than $t_{H}\sim10\,\mathrm{Gyr}$
\citep[e.g.][]{pre+10}. Scalar RR of $J$, on timescale $T_{RR}^{s}$
(red line) is shown for an assumed mean background stellar mass of
$\left\langle \protect\Ms\right\rangle =1\,\protect\Mo$ (top line),
and $\left\langle \protect\Ms\right\rangle =10\,\protect\Mo$ (bottom
line) (simplified by assuming a $J$-averaged $T_{c}^{GR}$, an early
approximation that is now superseded by the $\boldsymbol{\eta}$-formalism,
Sec. \ref{sss:eta}). $T_{RR}^{s}$ is shortest on the ${\cal O}(0.01\,\mathrm{pc})$
scale, where stellar mass BHs accumulate by mass segregation (Sec.
\ref{sss:SScusp}) and are eventually driven to orbital decay by the
emission of GWs (Sec. \ref{s:GRMBH}). Closer to the MBH, scalar RR
becomes inefficient due to quenching of RR by rapid prograde GR precession,
and further away by quenching due to retrograde mass precession (Sec.
\ref{ss:RR}). It is noteworthy that the red giants appear dynamically
relaxed \citep{gen+00}, despite the fact that a large fraction of
them are shorter lived than either stellar NR or scalar RR. This may
be a hint that additional relaxation mechanisms are at work (Sec.
\ref{sss:MPs}). Vector RR of the orbital plane, on timescale $T_{RR}^{v}$,
is much faster (magenta line), and could be the mechanism responsible
for the inner truncation of the observed stellar disk. Both scalar
and vector RR may explain the randomization of the S-stars (a cluster
of MS B stars) that orbit the inner $\mathrm{few\times}0.01\,\mathrm{pc}$
of the Galactic Center (Sec. \ref{ss:SgrA}).}
\end{figure}

This hierarchy of relaxation timescales can have observable consequences.
\textbf{Figure} \ref{f:GCrelax} shows the run with $a$ of the various
relaxation times in the Galactic Center, which suggests that RR plays
a role in establishing some of the systematic trends observed in the
different stellar populations there \citep{hop+06a}. 

Some of these theoretical insights on the nature of relaxation around
a MBH were verified by direct $N$-body simulations \citep[e.g.][]{eil+09}.
However, a full scale $N$-body simulation that spans the large dynamical
and temporal range in the singular potential of the MBH is extremely
challenging and still impractical. Even $N$-wire simulations \citep{tou+09},
which evolve in time the phase-averaged wires directly, remain computationally
expensive. One feasible method is to recast RR in terms of effective
diffusion coefficients, and use Monte-Carlo simulations to evolve
test stars in phase-space and statistically derive the stellar DF
(Sec.\ref{sss:eta}). 

\section{Dynamics of close MBH/star encounters}

\label{s:LC}

\subsection{The Newtonian loss cone}

\label{ss:NewtonLC}

A star on a nearly zero energy (parabolic) orbit $E=G\Mbh/2a\sim0$
(note stellar dynamical inverse sign convention $E>0$ for bound orbits)
has eccentricity $e\to1$, velocity $v(r)\simeq\sqrt{2G\Mbh/r}$,
specific angular momentum $J=\sqrt{G\Mbh a(1-e^{2})}=rv\sin\theta$
relative to the MBH and a Keplerian periapse $r_{p}\simeq J^{2}/2G\Mbh$.
The proximity to the MBH can lead to prompt stellar destruction if
$r_{p}$ lies inside the event horizon, or inside the tidal disruption
radius. A star at distance $r$ from the MBH has a periapse $\le r_{p}$
if its velocity vector $\boldsymbol{v}$ lies inside a cone centered
on $-\boldsymbol{r}$, with an opening angle of $\sin\theta\simeq\sqrt{r_{p}/r}$.
Generalizing to bound, non-zero energy orbits, the loss-cone is the
phase space volume of orbits with $J(E)\le J_{lc}(r_{d})\simeq\sqrt{\left(2-r_{d}/a\right)2G\Mbh r_{d}}$,
where $r_{d}$ is the maximal periapse for destruction. In terms of
the circular angular momentum $J_{c}=\sqrt{G\Mbh a}$, $j_{lc}=J_{lc}/J_{c}=\sqrt{(2-r_{d}/a)r_{d}/a}=\sqrt{1-e_{lc}^{2}}$.
(cf \textbf{Figure} \ref{f:EJschematic} left). The phase space volume
of the loss-cone is very small. For example, in a galactic nucleus
similar to the Milky Way ($\Mbh=4\times10^{6}\,\Mo$, $r_{h}\sim3\,\mathrm{pc}$),
the angular size of the loss-cone for the tidal disruption of $1\,\Mo$
MS stars at $r\sim r_{h}$ is $\theta_{lc}\sim\sqrt{r_{t}/r_{h}}\sim10^{-3}$
rad (Sec. \ref{s:tide}). This corresponds to a fraction of $\theta_{lc}^{2}/4$
of the stars at $r_{h}$, (assuming an isotropic stellar distribution),
or $Q\theta_{lc}^{2}/4\sim{\cal O}(1)$ stars on a loss-cone orbit.
Since such stars are promptly destroyed in less than an orbital period,
the steady-state rate of TDEs is set by the processes that repopulate
the loss-cone.

\subsubsection{$E$ vs $J$ diffusion}

\label{sss:EvsJ}

Diffusion into the loss-cone is characterized by a clear separation
of scales. It is faster to reach the MBH by diffusion in $J$ than
in $E$. This property plays a crucial role in simplifying the analysis
of loss cone dynamics (Sec. \ref{sss:eta}). In the impulsive limit,
2-body scatterings over a short time $\mathrm{d}t$ change only the
velocity of the test particle, but not its position, and the change
is isotropic and its rms small, $|\Delta\boldsymbol{v}|\ll v$. The
specific kinetic energy of the test star $E=v^{2}/2$ then changes
by $\Delta E\simeq\boldsymbol{v}\cdot\Delta\boldsymbol{v}$, and so
$\Delta E/E\sim\Delta a/a\sim{\cal O}(\Delta v/v)$. Similarly, the
specific angular momentum $J=|\boldsymbol{r}\times\boldsymbol{v}|$
changes by $\Delta J\simeq|\boldsymbol{r}\times\Delta\boldsymbol{v}|$,
and so $\Delta J/J_{c}\sim{\cal O}(\Delta v/v)$, where $J_{c}\sim rv$.
The relative (logarithmic) changes are therefore $\Delta\log E\sim\Delta\log a\sim{\cal O}(\Delta v/v)$
and $\Delta\log J\sim(\Delta J/J_{c})(J_{c}/J)\sim{\cal O}(\Delta v/v)/j$
(i.e., the diffusion ``velocity'' in $\log J$ is $1/j$ times faster
than in $\log E$). Orbits that are already somewhat eccentric evolve
so much faster in $J$ than $E$ that $E$ (and $a$) can be approximated
as constant. The relaxation time for diffusing into the loss cone
is $T_{J}=J^{2}/D_{JJ}\sim j^{2}T_{E}$ (\textbf{Figure} \ref{f:EJschematic}).
Moreover, the logarithmic distance to the innermost stable orbit in
the $a$ direction is longer than in the $J$ direction. The innermost
stable circular orbit (ISCO) of a non-spinning MBH is at $a_{\bullet}=6r_{g}$,
while the lowest stable angular momentum for a parabolic orbit (a
good approximation for plunge orbits) is $J_{\bullet}=4r_{g}c$. Therefore,
the logarithmic distances to the MBH from an initial orbit at $(a_{0},J_{0})$
are $\Delta_{a}=\log(a_{0}/6r_{g})$ and $\Delta_{J}=\frac{1}{2}\log(j_{0}^{2}a_{0}/16r_{g})$,
where $j_{0}=J_{0}/J_{c}(a_{0})$, and their ratio is $\Delta_{a}/\Delta_{J}=2\left[\log(a_{0}/r_{g})-\log6\right]/\left[\log(a_{0}/r_{g})-\log16+\log(j_{0}^{2})\right]>2$.
In particular, $\Delta_{a}/\Delta_{J}\to\infty$ as $j_{0}\to0$.
Since the velocity along the $J$-direction is faster while the distance
is shorter, diffusion into the loss-cone is primarily in angular momentum.

\subsubsection{Plunge vs inspiral}

\label{sss:PvsI}

There are in general two dynamical modes by which objects fall into
a MBH \citep{ale+03b}. One is by direct plunge, as discussed above,
where the object is promptly removed from the system once it approaches
the MBH closer than some destruction radius $r_{d}$. The removal
needn't involve actual destruction: for example, the tidal separation
of a binary removes the binary but does not destroy its constituent
stars (Sec. \ref{sss:TidalSeparation}). The distinguishing trait
of a plunge process is that the object is required to remain on the
plunging orbit only long enough to pass through periapse once, less
than one orbital time. Plunging orbits occur in one of two dynamical
regimes, defined by the ratio the orbital period $P(a)$ to the relaxation
time across the loss-cone, $T_{J}(a)=j_{lc}^{2}T_{E}(a)$: the empty
and full loss-cone regimes \citep{lig+77}. 

Close to the MBH, where $P(a)/T_{J}(a)<1$, the star plunges almost
unperturbed into the MBH, and the time-averaged phase-space density
in the loss-cone is nearly zero, since it takes slow diffusion to
supply new stars to plunge orbits. This is the empty, or diffusive,
loss-cone regime. This regime extends up to a critical sma $a_{p}$,
defined by $P(a_{p})/T_{J}(a_{p})=1$. The total plunge rate from
the empty loss-cone regime is approximately \citep{bar+16} 
\begin{equation}
\Gamma_{p}(a_{p})\approx\frac{5}{32}\frac{\Ns(<a_{p})}{\log[1/j_{lc}(a_{p})]T_{E}(a_{p})}=10\frac{\log Q}{\log[1/j_{lc}(a_{p})]}\frac{\Ns^{2}(<a_{p})}{Q^{2}P(a_{p})}\,,\label{e:PlungeRate}
\end{equation}
where $T_{E}$ is expressed in terms of the system parameters by Eq.
(\ref{e:Tss}), and where a single mass Bahcall-Wolf steady state
solution is assumed. 

Far from MBH, where $a>a_{p}$ and $P(a)/T_{J}(a)>1$, stars are scattered
in and out of the loss-cone many times before they reach periapse,
and therefore the loss-cone is full, and the stellar distribution
is effectively isotropic. In the full loss-cone regime, $\Gamma_{p}\sim j_{lc}^{2}\Ns/P$.
Analysis of the relative contributions to the plunge rate from the
empty and full loss-cone regimes around a MBH shows that most of the
plunging stars come from $a_{e}\sim\min(a_{p},r_{h})$, and that the
empty loss-cone rate Eq. (\ref{e:PlungeRate}) provides a reasonable
approximation for the total rate from both regimes with the substitution
$a_{p}\to a_{e}$ \citep{lig+77,sye+99}, which for $\Mbh\gtrsim10^{6}\,\Mo$,
where $a_{p}>r_{h}$, yields $\Gamma_{p}\sim{\cal O}(1/P_{h})$ (Eq.
\ref{e:TNR}).

The second mode of getting to the MBH is by inspiral, where the orbit
decays gradually by some dissipation mechanism that extracts a small
fraction of the orbital energy and angular momentum every orbit, until
the orbit shrinks below the ISCO. The dissipation mechanisms of interest,
GW emission (Sec. \ref{ss:EMRIs}), tidal heat dissipation (Sec. \ref{p:squeezar}),
or hydrodynamical interaction with a massive accretion disk (Sec.
\ref{s:disk}) are typically strongly decreasing functions of radius,
and so most of the orbital energy is extracted near periapse. 

In contrast to prompt plunge, inspiral down to the ISCO is gradual,
and takes many orbital times, during which the orbit is susceptible
to perturbations by the background stars. These can abort the inspiral
before the star reaches the MBH, either by deflecting it to a lower
eccentricity and larger periapse orbit, where dissipation is inefficient,
or by deflecting it directly to a plunge orbit. The total time for
inspiral increases with the initial sma. The race between inspiral
and orbital diffusion limits inspiral to stars that begin the process
close enough to the MBH so that the total inspiral time is faster
than the relaxation time. Phase space is therefore separated into
two regimes, which are approximately described by a critical sma for
inspiral, $a_{i}$ (\textbf{Figure} \ref{f:EJschematic}). At $a>a_{i}$,
stars that reach the MBH do it with high probability by direct plunge,
while at $a<a_{i}$ they do it with high probability by inspiraling
into it. The transition between the two regimes is quite sharp \citep{hop+05}.
There is no full loss-cone regime analogue for inspiral events. The
total inspiral rate can be approximated by \citep{ale+03b,bar+16}
\begin{equation}
\Gamma_{i}\approx\Gamma_{p}(a_{i})\sim(a_{i}/r_{h})\Gamma_{p}(r_{h})\,,\label{e:InspiralRate}
\end{equation}
where the last approximate relation neglects the logarithmic diffusion
terms and assumes $\alpha=7/4$. Because the inspiral time is much
longer than the time to plunge, $a_{i}\ll r_{h}$, and the inspiral
rate is much lower than the plunge rate. This simply reflects the
fact that there are many more stars inside $r_{h}$, which can diffuse
to plunge orbits, than there are stars inside $a_{i}$, which can
diffuse to inspiral orbits. For example, for a Milky Way-like nucleus,
where $r_{h}\sim\mathrm{few\times1}$ pc and $a_{i}\sim\mathrm{few\times0.01}$
pc \citep{hop+05}, the ratio of GW inspiral rate to plunge rate \footnote{\label{fn:NewtonianGW}This is considered in the Newtonian loss-cone
context, in spite of the inclusion of GW dissipation, since any type
of dissipation can be represented by the inspiral formalism, but the
non-dissipative dynamics are approximated as Newtonian (Sec. \ref{sss:NewtonRR}).} is $\Gamma_{i}/\Gamma_{p}\sim{\cal O}(0.01)$.

\subsubsection{Resonant Relaxation in the Newtonian limit}

\label{sss:NewtonRR}

Early studies of the loss-cone problem \citep{fra+76,lig+77,sha+78}
were confined to the Newtonian limit and did not include RR, which
was not yet discovered. Nevertheless, the consequences of this omission
for global loss-rate estimates turn out to be small, due to a coincidence
that can be fully explained only in the context of the relativistic
loss-cone (Sec. \ref{ss:relLC}). RR does not affect the plunge rates
much, since most plunges originate at $a\sim r_{h}$, where mass precession
quenches scalar RR (Sec. \ref{ss:RR}; \textbf{Figure} \ref{f:GCrelax}).
The relative contribution from the fewer stars closer to the MBH,
where RR is dynamically important, is small \citep{rau+96}. In contrast,
the branching ratio between inspiral and plunge events, if treated
in the Newtonian limit, appears to be dramatically (but incorrectly)
affected by RR. This is because in the absence of GR Schwarzschild
precession, the RR torques remain strong even on low-$J$ orbits,
and all stars evolve rapidly in $J$ and plunge into the MBH before
they can diffuse below $a_{i}$ and inspiral into the MBH \citep{hop+06a,bar+16}.
As a result, the inspiral rate drops to zero. It is only when both
the non-dissipative (precession) and dissipative (GW) GR terms are
included in the dynamics self-consistently, that the quenching effect
of GR precession is found to strongly limit the role of RR for the
loss rates (\textbf{Figure} \ref{f:EJschematic} right). The bottom
line is that the naive treatment of the loss-cone problem in the Newtonian
limit, which neglects both GR dynamics (apart for the GW dissipation)
and RR (in spite of the fact that it is a Newtonian process), yields
by coincidence the correct order of magnitude for the plunge and inspiral
rates (\textbf{Figure} \ref{f:LossRates}). 

\subsection{The relativistic loss cone}

\label{ss:relLC}

The extension of the essentially Newtonian treatment of loss-cone
dynamics (notwithstanding the GW dissipation term\textemdash see footnote
\ref{fn:NewtonianGW}) to the relativistic regime was made necessary
by the early realization that fast GR precession of an eccentric test
star will rapidly switch the sign of the residual torques exerted
on it by the nearly fixed background, and quench RR \citep{rau+96}.
This has important implications for EMRIs. \citet{hop+06a} conjectured
that GR precession could prevent RR from pushing all potential EMRI
sources into prompt plunge trajectories, because the ${\cal O}(\beta^{2}j^{-2})$
GR precession becomes significant before ${\cal O}(\beta^{5}j^{-7}Q^{-1})$
GW dissipation \citep{ale15}. This would decouple the sources from
the background and allow them to inspiral gradually and produce detectable
quasi-periodic GW signals. This departure from the simplifying assumption
that the test star is statistically similar to the background (Sec
\ref{ss:RR}) was treated in those early studies as equivalent to
assuming that the background is precessing randomly relative to the
test star, that is, GR precession was introduced as a stochastic perturbation
of the background stars \citep{rau+96}. 

\subsubsection{The Schwarzschild Barrier}

\label{sss:SB}

The first self-consistent post-Newtonian $N$-body simulations of
plunge and inspiral events \citep{mer+11} indicated however that
GR precession is not well described that way, and in particular, that
the coherent behavior of RR cannot be approximated by a Markov process.
Not only did GR precession quench RR before the stars entered the
GW-dominated regime, as conjectured, but the stellar trajectories
seemed to encounter a barrier in phase-space (the so-called ``Schwarzschild
Barrier'', SB) that prevented them from evolving to $j\to0$, and
plunging or inspiraling into the MBH (cf \textbf{Figure} \ref{f:SBAI}
left). Instead, the stars appeared to linger near the barrier for
roughly $T_{c}$, while their orbital parameters oscillated at the
Schwarzschild precession frequency. An early analysis suggested that
this behavior was related to precession under the influence of a residual
dipole-like force \citep{ale10,mer+11}. Larger scale $N$-body simulations
confirmed that GR quenches RR near the SB \citep{bre+14}. However,
the exact nature of the SB and the interpretation of stellar dynamics
near it remained controversial \citep[e.g.][]{ant+13}.

\begin{figure}
\includegraphics[width=0.95\textwidth]{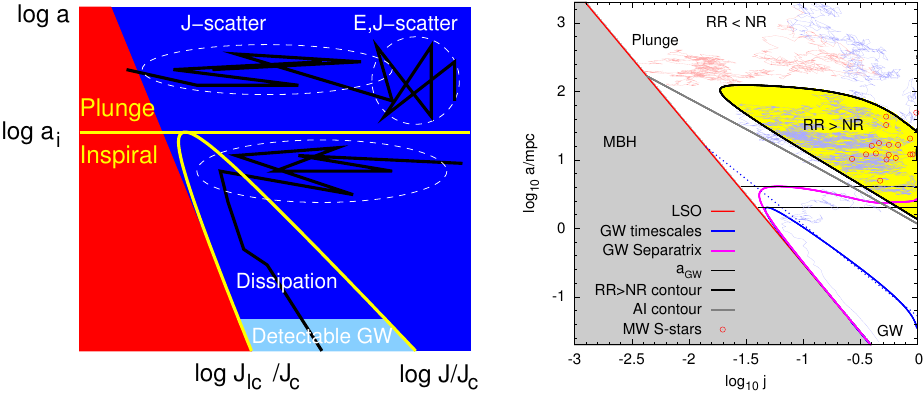}
\centering{}\caption{\label{f:EJschematic} \textbf{Left}: A schematic of loss-cone dynamics
in $(a,j)$ phase-space. At $J\lesssim J_{c}$ the scattering rates
in $\log a=-\log E$ and $\log J$ are comparable, but once $J\ll J_{c}$,
the scattering along $\log J$ is much faster (Sec. \ref{sss:EvsJ}).
Close to the MBH all random trajectories along $\log J$ cross the
GW line into a region where GW dissipation is faster than NR, and
therefore they inspiral into the MBH with high probability. Above
the tip of the GW line, at a critical sma $a_{i}$, all stars plunge
directly into the MBH with high probability. \textbf{Right}: An exact
calculation of the critical lines and regions, for a model of the
Milky Way with $Q=4\times10^{5}$, mass-precession coherence time
and Gaussian noise \citep{bar+16}. Orbits in the gray area below
the last stable orbit (red) are unstable and promptly plunge into
the MBH event horizon (Monte Carlo-generated plunge track example
in light red line, see footnote \ref{fn:MC}). Where RR diffusion
is faster than NR diffusion (yellow region), RR dominates the dynamics.
The S-stars observed near the MBH of the Milky Way (red circles) \citep{gil+09}
lie in the RR dominated region (Sec. \ref{ss:SgrA}, Sec \ref{ss:RR}).
Adiabatic invariance (AI) suppresses RR torquing below the AI line
(gray). Inside the phase-space region delimited by the GW line (blue),
GW dissipation is faster than NR $J$-scattering and orbits spiral
into the MBH by the emission of GW (Monte Carlo-generated inspiral
track example in light blue line). The critical sma for EMRIs, $a_{i}=a_{GW}$
(thin black line), corresponds to the maximum of the GW curve; below
it stars become EMRIs before they cross the last stable orbit line\@.
The approximate power-law GW line with the often assumed simplification
$j_{lc}\to0$ (dotted blue line), substantially over-estimates $a_{GW}$.
The exact separatrix streamline (magenta line) provides a more accurate
estimate of $a_{GW}$ than either of the timescale-based GW lines. }
\end{figure}

\begin{figure*}
\begin{raggedright}
\begin{tabular*}{0.33\textwidth}{@{\extracolsep{\fill}}c}
\includegraphics[width=0.95\textwidth]{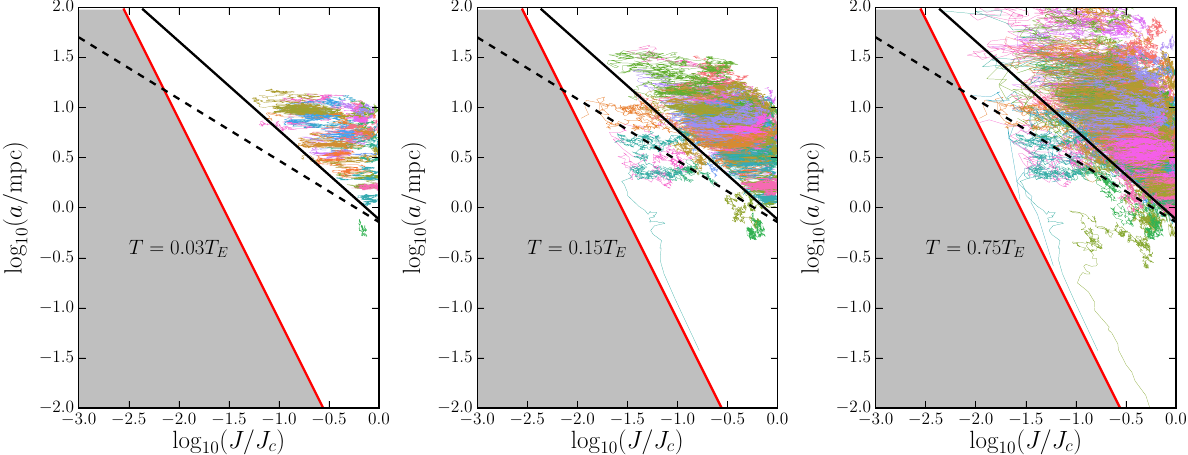}\tabularnewline
\end{tabular*}
\par\end{raggedright}
\caption{\label{f:SBAI} Snapshots of stellar tracks from a Monte Carlo integration
of the FP equation describing orbital evolution around a MBH under
the combined effects of NR, RR and GW emission (\citealp{bar+16};
see footnote \ref{fn:MC}). The snapshots are shown at increasing
fractions of $T_{E}$, the NR energy diffusion time. The phase space
is marked as in \textbf{Figure} \ref{f:EJschematic}, but for simplicity
only the AI locus (solid black line at $j_{AI}(a)=\sqrt{T_{c}(a)/P_{GR}^{0}(a)}$)
and the initially mis-identified SB locus of \citep{mer+11} (dashed
black line) are shown. It is now understood that the SB effect is
due to AI \citep{bar+14}. On short timescales (left panel), stellar
orbits evolve rapidly in angular momentum due to RR, but only above
the AI locus, since beneath, RR is strongly suppressed. However, on
longer timescales NR randomizes the orbits, the AI barrier is breached
and the phase space density approaches the maximal entropy solution
(right panel). }
\end{figure*}

\subsubsection{The $\boldsymbol{\eta}$-formalism}

\label{sss:eta}

The SB phenomenon lies in the difficult-to-treat interface between
deterministic Hamiltonian dynamics and stochastic kinetic theory.
The $\boldsymbol{\eta}$-formalism \citep{bar+14} provides a formal
framework for describing coherent relaxation and secular processes
around a MBH, and succeeds in explaining the SB phenomenology that
is observed in $N$-body simulations (\textbf{Figure} \ref{f:SBAI}),
and in reproducing the $N$-body loss-rates \citep{bar+16}. 

The key idea of the $\boldsymbol{\eta}$-formalism is that the effect
of the background stars on the test orbit can be represented as a
time-correlated noise model\footnote{The noise can be approximated as independent of $J$ (i.e. fully correlated
in $J$), because the mean free path in $J$ is small: $\tau_{N}T_{c}^{M}\ll J_{c}$
(Sec. \ref{ss:RR}) and $\tau_{N}T_{c}^{SQ}<J_{c}$ \citep{koc+15}.}. A perturbative expansion of the phase-averaged post-Newtonian Hamiltonian
reveals that to leading order, the noise is a vector in angular momentum
phase space, $\boldsymbol{\eta}(t)$. The noise model is characterized
by an auto-correlation function with a coherence time scale $T_{c}$.
Stochastic equations of motion are then derived from the Hamiltonian
and used, together with random realizations of correlated noise, to
evolve test orbits in phase space. The resulting dynamics are found
to depend critically on the temporal smoothness (degree of differentiability)
of the noise. When the noise is smooth (infinitely differentiable,
as expected for noise generated by the superposition of many smooth
background orbits), its power-spectrum drops fast beyond some maximal
frequency. A star precessing faster than that cutoff frequency is
effectively decoupled from the perturbing background by adiabatic
invariance (AI). For GR precession, the AI/SB barrier in phase space
is the locus\footnote{Eq. \ref{e:jAI} \citep{bar+14} corrects the SB's misidentified locus
of \citet{mer+11}.}
\begin{equation}
j_{AI}(a)\simeq\sqrt{T_{c}(a)/P_{GR}^{0}(a)}\,,\label{e:jAI}
\end{equation}
where $P_{GR}^{0}=(a/3r_{g})P$ is the GR precession period for $j=1$.

Although the noise is correlated, it is possible derive (and validate
against results from the stochastic equations of motion) effective
diffusion coefficients\footnote{General maximal entropy considerations fix the steady state $J$-DF
for $J$-evolution under RR. The FP equation then imposes a relation
(the fluctuation\textendash dissipation relation, \citealp{cal+51})
that must be satisfied by any valid RR diffusion coefficients (see
also footnote \ref{fn:0current}).}, which allow to evolve the probability density function $\rho(j)$
with the FP equation\footnote{$\partial\rho/\partial t=\nicefrac{1}{2}\nicefrac{\partial}{\partial j}\left\{ jD_{jj}\nicefrac{\partial}{\partial j}(\rho/j)\right\} $
with a parametric drift coefficient $D_{j}=\nicefrac{1}{2j}\nicefrac{\partial}{\partial j}(jD_{jj})$.}. The RR diffusion coefficient $D_{jj}^{RR}(j;a)$ is proportional
to the power-spectrum of the noise at the precession frequency $\nu_{GR}(j;a)=2\pi/P_{GR}^{0}(a)j^{2}$,
and therefore for smooth noise models, AI is manifested as an extremely
steep fall in the RR diffusion rate for $j<j_{AI}(a)$. The AI/SB
barrier is not a reflecting barrier, but a locus in phase space beyond
which only NR diffusion is effective. RR diffusion beyond the AI/SB
barrier does not fall to zero identically. Since diffusion to yet
lower $j$ slows further down, while diffusion to higher j speeds
further up, orbits statistically seem to bounce away from the SB back
to higher $j$. The highly suppressed but non-zero RR diffusion beyond
the barrier means that the zero-density front advances to lower $j$
logarithmically slowly in time. However, for practical purposes the
barrier can be considered fixed. 

The ability to implement the $\boldsymbol{\eta}$-formalism of RR
as diffusion is of great practical value. It allows to model the dynamics
of galactic nuclei in the $\Ns\to\infty$ limit by evolving the probability
density by Monte-Carlo simulations of the combined NR and RR diffusion,
including additional processes such as GW decay\footnote{\label{fn:MC}The NR+RR+GW FP equation is evolved in time by Monte
Carlo by incrementing the test star's phase space position in dimensionless
energy and angular momentum $(\epsilon,j)$ over small time interval
$\mathrm{d}t$ using the NR and RR diffusion coefficients and approximations
for the energy and angular momentum dissipation by GW over $\mathrm{d}t$,
$\mathrm{d}\epsilon^{GW}$ and $\mathrm{d}j^{GW}$: $\mathrm{d}\epsilon=D_{\epsilon}^{NR}\mathrm{d}t+\gamma_{1}\sqrt{D_{\epsilon\epsilon}^{NR}\mathrm{d}t}+\mathrm{d}\epsilon^{GW}$
and $\mathrm{d}j=D_{j}^{NR}\mathrm{d}t+\gamma_{2}\sqrt{D_{jj}^{NR}\mathrm{d}t}+D_{j}^{RR}\mathrm{d}t+\gamma_{3}\sqrt{D_{jj}^{RR}\mathrm{d}t}+\mathrm{d}j^{GW}$,
where $\gamma_{1}$, $\gamma_{2}$ ,and $\gamma_{3}$ are randomly
drawn normal variates, with $\gamma_{1}$ and $\gamma_{2}$ correlated
with a correlation coefficient $\xi=D_{\epsilon j}/\sqrt{D_{\epsilon\epsilon}D_{jj}}$. } \citep{bar+16}. 

\subsubsection{Steady state phase space structure and loss rates}

\label{sss:LCSteadyState}

The $\boldsymbol{\eta}$-formalism does not take into account diffusion
by NR, which is unaffected by precession (Sec. \ref{ss:GenRelax}),
and continues independently across phase space. \textbf{Figure} (\ref{f:EJschematic}
right) shows the region in phase space where the RR diffusion rate
is faster than NR, and the locus of the AI line, below which the RR
diffusion rate is essentially zero. The RR-dominated region is separated
from the loss-lines (plunge and GW inspiral), and therefore the bottle-neck
for the loss rates remains slow NR, with only a modest boost from
RR (Sec. \ref{sss:NewtonRR}). On timescales ${\cal O}(T_{E})$, NR
erases the density drop beyond the AI/SB barrier, and the system rapidly
approaches the maximal entropy configuration (\textbf{Figure} \ref{f:SBAI}),
as it should, irrespective of the nature of the randomization process.
The existence of an RR-dominated region in phase space may however
play an important role in the orbital evolution of special populations
in galactic nuclei, e.g. the S-stars in the Galactic Center (Sec.
\ref{ss:SgrA}, Sec. \ref{sss:TidalSeparation}). 

\begin{figure}
\begin{centering}
\includegraphics[width=0.6\columnwidth]{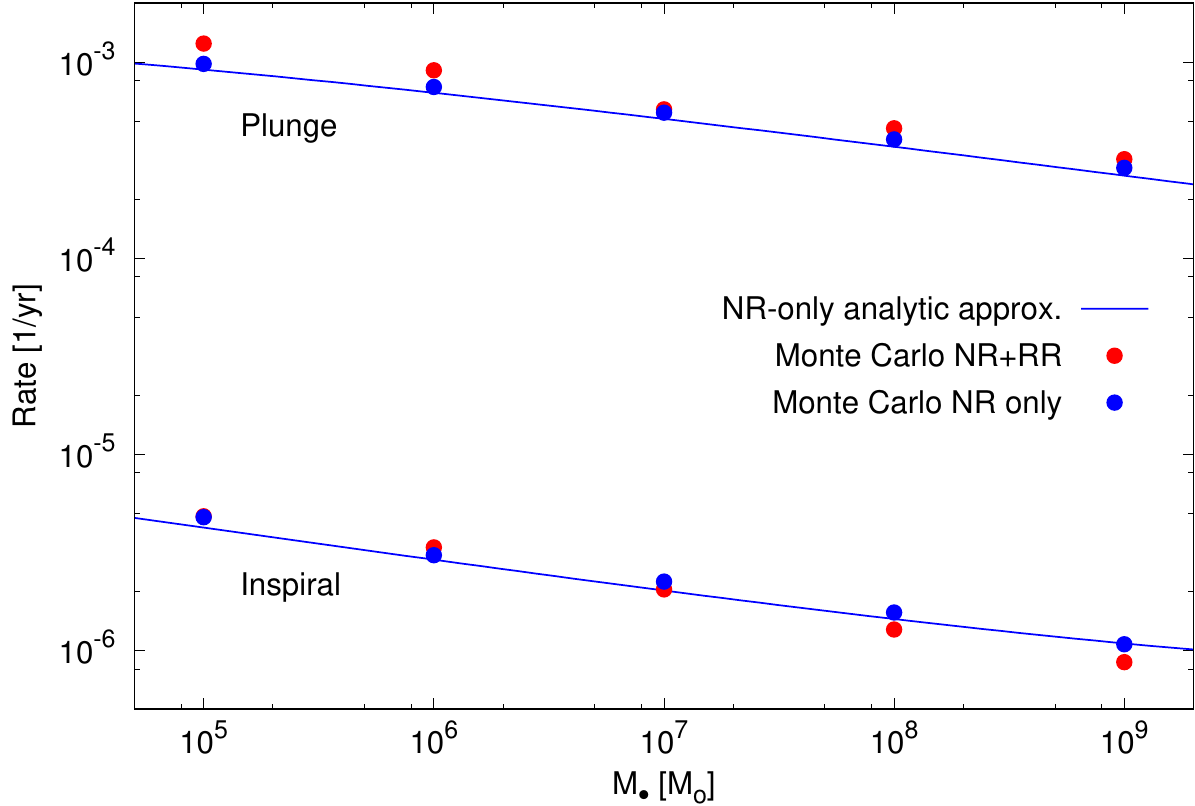}
\par\end{centering}
\caption{\label{f:LossRates}Plunge and EMRI rates per galaxy in steady state,
as function of $\protect\Mbh$, assuming a simplified single mass
population composed of $10\,\protect\Mo$ stars (``stellar BHs''),
scaled to the Galactic Center by the $\protect\Mbh/\sigma$ relation
with $\beta=4$. The rates were estimated by Monte Carlo simulations
that implement the $\boldsymbol{\eta}$-formalism (Sec. \ref{sss:eta})
(circles), and are compared to NR-only analytic approximations (lines)
(adapted from \citealt{bar+16}). The analytic estimates match the
corresponding NR-only Monte Carlo simulations well, and deviate only
slightly from the full NR+RR simulations (Sec. \ref{sss:LCSteadyState}).
The plunge rates are evaluated in the empty loss-cone limit, which
gradually over-estimates the true rate below $\protect\Mbh\lesssim10^{6}\,\protect\Mo$,
where $a_{p}<r_{h}$ (Sec. \ref{sss:PvsI}). }
\end{figure}

\textbf{Figure} (\ref{f:LossRates}) shows the plunge and inspiral
rates per galaxy, as function of $\Mbh$, that were estimated by Monte
Carlo simulations that used NR and effective RR diffusion coefficients
to evolve test stars in phase space ($\Mbh/\sigma$ with $\beta=4$
assumed). The results confirm and calibrate the weak approximate $\Mbh^{-1/4}$
dependence of the rates \citep{hop09b}. The close similarity between
the rates with and without RR shows that the strong suppression of
RR below the AI line indeed decouples the EMRIs from the background,
and allows them to proceed unimpeded, as conjectured \citep{hop+06a}.

\subsubsection{Other approaches to modeling dynamics near a MBH}

Dynamics near a MBH involve many processes, including NR, RR and secular
Newtonian and GR processes. Direct relativistic $N$-body simulations
(with either full or perturbative GR) \citep{mer+11,bre+14} generate
by construction all these processes. However, the interpretation of
the results is very difficult since these complex dynamics are entangled.
Moreover, special regularization techniques are required for maintaining
high numeric accuracy over the large dynamical range in the diverging
potential of the central mass \citep[e.g.][]{mik+06}. These are very
expensive computationally, and limit the simulations to unrealistically
small $\Ns$, which generally cannot be scaled up to astrophysically
relevant values since different dynamical mechanisms scale differently
with $\Ns$ \citep[e.g.][]{heg+03}.

Since most of the questions of interest in the modeling of processes
near MBH involve integration over timescales much longer than an orbital
period, one alternative to direct $N$-body simulations is ``$N$-wire''
simulations \citep{tou+09}, which treats the dynamics directly in
terms of the orbit-averaged mass wires (Sec. \ref{ss:RR}) with timesteps
$>P$. Unfortunately, the computational load of each timestep is currently
still too heavy to provide clear advantage over direct $N$-body simulations.

A recent attempt to derive a self-consistent description of the background
noise and its correlations from the Balescu-Lenard master equation,
may ultimately provide an \emph{ab-initio} description of the noise
\citep{fou+16}. However, the feasibility of its adaptation to practical
applications remains to be proven. 

\subsection{Non-collisional loss-cone refilling}

The angular momentum $\boldsymbol{J}$ is an integral of motion in
spherical potentials, and therefore collisions (NR or RR) are necessary
for refilling the loss cone with stars on $J<J_{lc}$ orbits. In non-spherical
potentials this is no longer the case, and the angular momentum of
individual orbits can change over time, whether in a restricted $J$-range
or in an unrestricted one that includes  $J=0$ (e.g. chaotic orbits
in a triaxial potential). When the range includes $J\to0$, the orbits
are called centrophilic. Stars on such families of orbits can be driven
into the loss-cone by collisionless torques, until the entire reservoir
is ``drained'' (\citealt{vas+13}; \citealp[See review by][]{vas14}).
The time for collisionless torquing to push a star into the loss-cone
is generally much shorter than NR diffusion, and as long the supply
of centrophilic orbits holds, the loss-rate reaches, or even surpasses
the maximal rate of collisional loss-cone refilling, which is given
by the full loss-cone rate (Sec. \ref{sss:PvsI}). Even after the
reservoir is drained, the lowered symmetries of galactic potentials
that allow collisionless torquing of centrophilic orbits effectively
increase the loss-cone, since slow relaxation now needs only to refill
the larger phase-space of centrophilic orbits, and the collisionless
torques will rapidly take the stars on to the loss-cone.

This is what is found for axisymmetric galaxies \citep{mag+99}, where
the draining phase is cosmologically brief and can therefore be neglected.
The lowered symmetry indeed expands the loss-cone into a larger ``loss-wedge'',
but the steady state increase in the plunge rate is only $\sim\times2$.
The situation is different for triaxial galaxies, where the draining
time can be longer than $t_{H}$, and the loss-cone is not in steady
state. In that case, the question whether collisionless torquing dominates
loss-cone dynamics directly reflects assumptions about the initial
conditions: what was the initial mix of orbital families and what
was the initial distribution of angular momentum. An additional uncertainty
is due to orbital scattering by the MBH, which may drive the stellar
potential near it to isotropy \citep{ger+85}. The role of non-collisional
loss-cone refilling in the cosmic loss rates remains unclear.

\section{Tidal interactions between stars and a MBH}

\label{s:tide}

A disruptive tidal interaction occurs when a star (or binary, or gas
cloud) of mass $\Ms$ and radius $\Rs$ approaches the MBH on an orbit
with periapse $r_{p}<r_{t}=Q^{1/3}\Rs$. The work done by the tidal
field transfers energy and angular momentum from the orbit to the
star and unbinds it. In the case of a non-disruptive impulsive tidal
interaction, such as a hyperbolic flyby with $r_{p}\gtrsim r_{t}$,
or an eccentric periodic orbit with $a\gg r_{p}\gtrsim r_{t}$, the
strong tides excite oscillations in the star, and possibly also lead
to some mass loss, but the star survives (Sec. \ref{p:squeezar}).

\paragraph*{Classification of tidal disruptions}

The characteristics of a tidal interaction with a BH, and the physical
processes that are required to describe it, depend on three length-scales:
The stellar radius $\Rs$, the tidal radius $r_{t}$ and the gravitational
radius of the BH, $r_{g}$. Their ratios can be expressed in terms
of the stellar break-up velocity, $\vs$, which measures the star's
self-gravity, and in hydrostatic equilibrium, also its pressure: $r_{t}/\Rs=Q^{1/3}$,
$r_{t}/r_{g}=(\vs/c)^{-2}Q^{-2/3}$ and $r_{g}/\Rs=(\vs/c)^{2}Q$.
There are generally five full disruption regimes, ordered by the mass
ratio $Q$ \citep{ale05}. 
\begin{enumerate}
\item $Q\ll1$ ($r_{g}\ll r_{t}\ll\Rs$): A weak Newtonian tidal interaction,
where the star's self-gravity and pressure dominate. Relevant when
a stellar BH is swallowed by a star, and can result in an exotic star
powered by accretion \citep[e.g.][]{tho+75}.
\item $Q\sim1$ ($r_{g}\ll r_{t}\sim\Rs$): A strong Newtonian tidal interaction
with significant mass loss and possible disruption, such as occurs
in a close interaction between a stellar BH and a massive star (Sec.
\ref{sss:exchange}). 
\item $Q\sim(c/\vs)^{2}$ ($\Rs\sim r_{g}<r_{t}$): A complete disruption
in the Newtonian regime, as would be the case for disruption by an
IMBH. 
\item $(c/\vs)^{2}<Q<(c/\vs)^{3}$ ($\Rs\ll r_{g}<r_{t}$): A complete tidal
disruption by a lower-mass MBH of the type considered here (e.g. $\SgrA$),
which can be treated as Newtonian to a good approximation\footnote{The subsequent circularization and accretion of the debris are in
the relativistic regime.}. 
\item $Q>(c/\vs)^{3}$ ($\Rs\ll r_{t}\ll r_{g}$): Tidal disruption inside
the event horizon. The star plunges into the MBH as a point particle
on a GR trajectory. 
\end{enumerate}

\subsection{Stellar disruption}

\label{ss:TDEs}

TDEs can be an important mass supply channel for lower mass MBHs.
This is seen by comparing the plunge rate $\Gamma_{t}\propto\Mbh^{-1/4}$
(Figure \ref{f:LossRates}) to the MBH mass, approximating a constant
rate over $t_{H}$ and $\Ms\sim1\,\Mo$, which yields $\Mbh\sim\Ms\Gamma_{t}t_{H}$
for $\Mbh\lesssim10^{7}\,\Mo$ (simulations indicate that the TDE
contribution is $(0.15-0.65)\Mbh$, \citealp{mur+91,fre+02}). For
such lower-mass MBHs $r_{t}>\rbh$ and therefore the accretion of
stars is luminous. Tidal accretion flares can signal the presence
of an otherwise quiescent growing MBH and probe accretion physics.

\paragraph*{Physics of stellar disruption}

The disrupted star is typically scattered into the loss-cone from
a long-period orbit with $a\sim r_{h}$ (Sec. \ref{sss:PvsI}), and
is therefore on a hyperbolic orbit relative to the MBH, with specific
orbital energy $\epsilon$ that is a small fraction of its binding
energy $\epsilon_{\star}=-\vs{}^{2}$, $\epsilon\sim{\cal O}(0.1)|\epsilon_{\star}|$
\citep[e.g.][]{ale+01b}. The orbit can therefore be approximated
as parabolic ($\epsilon=0$). The work done by the tidal field on
the star transfers orbital energy $\Delta\epsilon\sim G\Mbh\Rs/r_{t}^{2}\sim Q^{1/3}\epsilon_{\star}\gg\epsilon_{\star}$
to it, thereby disrupting the star\footnote{It was recently realized that $\Delta\epsilon$ hardly depends on
the penetration depth $b=r_{p}/r_{t}$, since once the star crosses
$r_{t}$, its self-gravity is negligible and the tidal field no longer
performs work against it \citep{sar+10,sto+13,gui+13}.}. Roughly half the stellar debris is ejected to infinity with positive
orbital energy up to $\epsilon_{\mathrm{out}}\sim\epsilon_{\star}+\Delta\epsilon\sim+\Delta\epsilon$,
while the other half is captured with negative orbital energy down
to $\epsilon_{\mathrm{in}}\sim\epsilon_{\star}-\Delta\epsilon\sim-\Delta\epsilon$.
The most bound debris stream returns to the point of disruption near
the MBH on an eccentric orbit after time $t_{\min}=(\pi/\sqrt{2})Q^{1/2}t_{\star}$.
The rest of the bound mass follows with a decreasing fall-back rate
$\dot{M}_{fb}=\Ms/(3t_{\min})(t/t_{\min})^{-5/3}$ \citep{ree88}\footnote{Note that this relation originally appeared in \citet{ree88} with
a known typo ($t^{-4/3}$).}. The peak fall-back rate $\max\dot{M}_{fb}=\Ms/(3t_{\min})$ exceeds
the Eddington limit for $\Mbh\lesssim3\times10^{7}\,\Mo$. Note that
the connection between the mass fall-back rate and the observable
accretion emission is neither direct nor obvious, because the gas
must first circularize before luminous accretion can proceed, because
it is not clear how the initial super-Eddington phase appears observationally,
and because the luminosity emitted at specific bands at different
times is not simply proportional to $\dot{M}_{fb}$. Though theoretically
unjustified, the $F_{\nu}\propto t^{-5/3}$ flux decline remains a
conventional criterion for distinguishing tidal flares from other
types of AGN activity.

This simplified initial picture of TDEs was subsequently refined and
modified. It is now understood that the early phases of the fall-back
(the rise and the settling on the asymptotic $t^{-5/3}$ decline)
depend on the details of the stellar structure \citep{lod+09}. Partial
tidal disruption leads to a faster than $t^{-5/3}$ decline, due to
the fact that some of the tidal interaction's energy is transferred
to the bound remnant at the expense of the debris \citep{gui+13}.
GR precession and frame-dragging appear to play a crucial role in
the post-disruption circularization of the debris streams, especially
if the orbit is bound \citep{dai+13}, by misaligning debris streams
of different energies and enhancing the shocks as they intersect near
the disruption point \citep{gui+15,bon+16b,hay+16}. GR Precession
may also link the penetration depth with the flare temperature, since
deeper penetration leads to a larger precession angle and an intersection
point deeper in the MBH potential \citep{dai+15}. 

\paragraph*{Observed tidal disruption flares}

More than 20 TDEs have been observed to date in the radio, optical,
UV, X-ray and $\gamma$-ray. This translates to a rate of $\sim10^{-5}\,\mathrm{yr^{-1}\,gal^{-1}}$
\citep{sto+16}. The predicted plunge rates shown in \textbf{Figure}
\ref{f:LossRates} ($\sim5\times10^{-4}\,\mathrm{yr^{-1}\,gal^{-1}}$
for $\Mbh=10^{7}\,\Mo$, \citealp{bar+16}) are in line\footnote{These plunge rates \citep{bar+16} are for horizon crossing, not tidal
disruption. The adaptation of the calculations to tidal disruption
involves substituting $J_{lc}=4r_{g}c$ by $J_{t}=\sqrt{2G\Mbh r_{t}}$
for an $e=1$ marginal tidal disruption orbit. However, since $J_{t}/J_{lc}=\sqrt{r_{t}/8r_{g}}\simeq2.4(\Mbh/10^{6}\,\Mo)^{-1/3}\sim{\cal O}(1)$
(Solar type star assumed), and since the plunge rate is a weak function
of $j_{lc}\ll1$ (Eq. \ref{e:PlungeRate}), the horizon crossing rates
are also reasonable estimates for the MS TDE rates. } with other predictions of $\Gamma_{t}\gtrsim10^{-4}\,\mathrm{yr^{-1}\,gal^{-1}}$
\citep{mag+98,wan+04}, and they all are systematically above the
observed rate by $\sim\times10$ \citep{sto+16}. The meaning of this
tension between the uncertain predicted rates and the observed ones
is still unclear.

Recent observations indicate that TDEs are over-represented in rare
galaxies that exhibit signs of a recently-ended ($0.1$\textendash $1$
Gyr old) intense star formation epoch \citep{arc+14}, as evidenced
by the strong presence of A-type stars\footnote{Of a TDE sample of 8 events, 3 are in A+E galaxies ($2\times10^{-3}$
of all galaxies), and 6 in a wider class of hosts ($2.3\times10^{-2}$
of all galaxies) that includes also quiescent galaxies with strong
Balmer absorption lines, a signature of A stars, \citealt{fre+16}.}. TDEs are over-abundant in these post-starburst galaxies by factors
of 33\textendash 190, and the implied TDE rate for this special class
of hosts is $10^{-3}\,\mathrm{yr^{-1}\,gal^{-1}}$ \citep{sto+16b}.
This further exacerbates the discrepancy between the predicted and
observed rates in all other galaxy types.

There are various possible explanations for the connection between
post-starburst galaxies and elevated TDE rates. If the starburst is
triggered by a recent galactic merger, a binary MBH in the center
may increase the TDE rate by several orders of magnitude \citep{che+11}.
However this phase is too brief to explain more than ${\cal O}(0.01)$
of the overall high TDE rate \citep{weg+11} and furthermore it requires
a coincidence between the decay times of binary MBHs and the lifespan
of A stars. Alternatively, the high TDE rate may follow a dissipative
flow of gas to the galactic nucleus that results in a central starburst
and an unusually high central stellar density. This leads to a high
TDE rate (if in steady state, $\Gamma_{t}\propto\Ns^{2}$, Eqs. \ref{e:Tss},
\ref{e:PlungeRate}). Observations of nearby A+E galaxy NGC~3156
indeed indicate a dense nucleus that can support a high TDE rate of
$\Gamma_{t}\sim10^{-3}\,\mathrm{yr^{-1}}$ \citep{sto+16}.

\paragraph*{Tidal detonation}

\label{p:detonation}

During a deep tidal encounter ($r_{p}/r_{t}\ll1$), the ballistic
trajectories of different mass elements in the disrupting star converge
(``pancake'') to the orbital plane of the star's center of mass,
and reach a transient state of high density, pressure and temperature
at periapse \citep{car+82}. The maximal compression and the temperature
depends sensitively on the hydrodynamics of the flow, which become
relevant once the temperature rises enough that the ballistic speed
becomes subsonic. Detailed simulations \citep{lag+93} indicate that
when $r_{p}/r_{t}\lesssim$${\cal O}(0.1)$, the density and temperature
reach high enough values, for a long enough time, to ignite runaway
nuclear fusion in the stellar core material, if it is already evolved
(or the star is a WD, \citealp{ros+09}) and composed of fast-fusing
heavier elements \citep[e.g.][]{ale05}. GR precession of the gas
streams further enhances the compression by leading to multiple density
maxima. The unusually energetic radio/X-ray source Sgr A East in the
Galactic Center may be the outcome of a tidal nuclear detonation event
\citep{dea+05}. 

\paragraph*{Tidal heating and inspiral}

\label{p:squeezar}

When a star is scattered to an eccentric orbit with periapse $r_{t}<r_{p}\lesssim2r_{t}$,
non-destructive tidal interactions on successive peri-passages transfer
orbital energy to internal oscillations, mostly at the fundamental
frequency $\nu_{\star}\sim\sqrt{G\Mbh/\Rs^{3}}\gtrsim\sqrt{G\Mbh/r_{p}^{3}}$
\citep{pre+77}. These cascade by non-linear mode coupling to higher
frequencies and ultimately dissipate as heat\footnote{In the absence of efficient dissipation, orbit/oscillations resonances
and/or reverse energy and angular momentum transfer back to the orbit,
are possible \citep{nov+92}. }. Possible orbit-oscillation runaway resonances are further suppressed
by the random phase perturbations due to the stellar background \citep{ale+03a}.
As the orbit decays and the star inspirals in, the heating rate grows
until the star approaches its Eddington luminosity. This tidally powered
transient source, a ``squeezar'' \citep{ale+03a} is ultimately
destroyed after ${\cal O}(10^{5}\,\mathrm{yr})$, either by radiatively
evaporating itself (a ``hot squeezar'' that radiates the tidal heat),
or by expanding beyond its Roche lobe until it is tidally destroyed
(an adiabatic ``cold squeezar''). The mean inspiral event rate by
tidal heating is much smaller than the TDE rate, $\sim0.05\Gamma_{t}$
(Sec. \ref{sss:PvsI}); the estimated mean number of squeezars around
an $\SgrA$-like MBH is $\sim0.1$\textendash $1$.

\subsection{Near misses}

The TDE rate is also roughly the rate of near misses ($r_{t}<r_{p}\lesssim\mathrm{few}\times r_{t}$),
since the cross-section for an encounter with periapse $\le r_{p}$
scales as $r_{p}$ due to gravitational focusing. A non-negligible
fraction of the stars inside the radius of influence of a lower-mass
MBH (such as $\SgrA$) have undergone a near-miss flyby. This is because
TDEs provide a substantial fraction of the mass for such MBHs (Sec.
\ref{ss:TDEs}), and the stellar mass inside $a\lesssim r_{h}$, where
most plunges originate, is also ${\cal O}(\Mbh)$. A second consecutive
non-destructive flyby is unlikely. Since the star's sma is typically
$a\sim r_{h}$ and its orbit lies between the empty and full loss-cone
regimes (Sec. \ref{sss:PvsI}), it will likely be deflected off the
plunging orbit by background stellar perturbations, or miss the MBH
due to its Brownian motion \citep{bah+77,cha+02c,mer+07} relative
to the cusp's center of mass \citep{ale+01b}. Furthermore, since
mass tidally lost from the stellar envelope and bound to the MBH adds
positive energy to the surviving star, the new apoapse could lie well
beyond $r_{h}$ \citep{man+13}. Near-misses can be important if a
single strong tidal interaction with the MBH has long-lasting effects
on the star. 

\paragraph*{Tidal scattering }

A strong non-destructive tidal interaction with an MBH leads to extreme
distortion, spin-up, mixing and mass loss. Of these, high spin and
mass-loss have the longest-lasting effects. Shortly after flyby, on
the thermal relaxation timescale, the star will be redder \citep{man+13}.
The long-term effect of high spin is hotter, more luminous stars with
abundance anomalies due to mixing. Such unusual populations are observed
in the Galactic Center \citep{car+00}. Quantitative estimates predict
that  ${\cal O}(0.01)$ of the stars in the radius of influence of
the Galactic Center have undergone a strong tidal scattering event
\citep{ale+01b,man+13}.

\paragraph*{Tidal disruption and tidal stripping of red giants}

Tidal interactions of red giants with a MBH differ from those of MS
stars due to the core / envelope dichotomy of red giant structure.
This enables near-miss flybys to tidally strip a substantial fraction
of the extended envelope while preserving the burning core. Due to
the large size of the giant, the fallback time is long, and the fallback
rate, and hence the luminosity, are lower than for a MS star \citep{mac+12}.

The gas streams ejected from a tidally disrupted object move initially
on ballistic orbits through the thin gaseous medium surrounding the
MBH. The debris streams from stripped red giants are less dense than
those from MS stars, both because the giant envelope is thinner, and
because the larger tidal radius corresponds to more spread-out streams.
The debris is therefore more susceptible to dissolution by the Kelvin-Helmholtz
instability as it orbits through the inter-stellar medium. This can
further decrease, or even completely throttle the fall-back rate on
the MBH and decrease the tidal flare luminosity \citep{bon+16}. It
thus appears unlikely that quiescent MBHs with mass $\Mbh>10^{8}\,\Mo$,
which only disrupt giants, will reveal themselves by tidal flares. 

The cloud G2, which is tidally interacting with $\SgrA$ (\citealp{gil+12};
Sec. \ref{ss:SgrA}), may be a clump in a debris stream of a recent
tidal stripping event \citep{gui+14}.

\subsection{Three-body exchange interactions}

\label{ss:3Body}

There are two variants of three-body exchange interactions with the
MBH, which involve two stars and the gravitational separation of a
bound 2-body system. One is the interaction of an incoming stellar
binary on a nearly radial unbound orbit with the tidal field of the
MBH. Such interactions can lead to a separation of the binary, where
one of its stars becomes bound to the MBH, and the other is ejected
as a single star (Sec. \ref{sss:TidalSeparation}). The other possible
exchange interaction occurs when a single incoming star on a nearly
radial unbound orbit interacts with a binary system composed of the
MBH and one of the stars closely bound to it. Such an interaction
can also lead to an exchange, where the incoming star ejects the bound
star, and takes its place as a bound companion of the MBH (Sec. \ref{sss:exchange}). 

\subsubsection{Tidal separation of a binary}

\label{sss:TidalSeparation}

The separation of a binary by the tidal field of a MBH is a mechanism
that can accelerate stars to velocities $\gg\vs$ \citep{hil88}\footnote{It is impossible to gravitationally accelerate a star to $\gg\vs$
by non-destructive interaction with a perturbing mass, unless the
perturber's own escape velocity is $\gg\vs$, which implies an MBH
(disruption cases 3 or 4, Sec. \ref{s:tide}). This limits the possibilities
to the interactions of a binary with a single MBH \citep{hil88},
or of a single star with a binary MBH \citep{yuq+03}.}. Such hyper-velocity stars (HVSs) can gain velocities well above
the escape velocity from their galaxy of origin, and thereby provide
evidence for the existence of a MBH (or a binary MBH) in the galactic
nucleus, provide information on the stellar population and the dynamics
near the MBH, and probe the galactic potential. 

\paragraph*{Dynamics of tidal separation by a MBH}

A binary of mass $M_{12}=M_{1}+M_{2}$ and sma $a_{12}$ that approaches
the MBH on a parabolic orbit closer than $r_{t}=Q^{1/3}a_{12}$ (here
$Q=\Mbh/M_{12}$) is separated by the tidal field. The tidal work
on the binary is $\Delta\epsilon_{12}\sim Q^{1/3}\epsilon_{12}$,
where $\epsilon_{12}=v_{12}^{2}=GM_{12}/a_{12}$ is the relative velocity
of the binary members. The radius of separation becomes the periapse
of the captured star (star 1), $r_{p1}=r_{t}=a_{1}(1-e_{1})$. The
capture sma is $a_{1}\sim G\Mbh M_{1}/2M_{12}\Delta\epsilon_{12}\sim Q^{2/3}a_{12}/4$
(neglecting $\epsilon_{12}\ll\Delta\epsilon_{12}$ and for $M_{1}=M_{2}$)
and therefore $e_{1}\sim1-Q^{-1/3}\sim1$. The ejected star (star
2) acquires a velocity at infinity (neglecting the potential of the
galaxy) $v_{2}\sim\sqrt{2(M_{12}/M_{2})\Delta\epsilon_{12}}\sim2Q^{1/6}v_{12}$.
For example, the capture of a $\Ms=10\,\Mo$, $\Rs=4\,\Ro$ S-star
in the Galactic Center by the separation of an equal mass binary with
$a_{12}=1\,\mathrm{AU}\simeq54\Rs$ would result an initial eccentricity
$e_{1}\simeq0.98$ and sma $a_{1}\simeq0.004\,\mathrm{pc}$ ($P\simeq12.5\,\mathrm{yr}$).
The ejection velocity would be $v_{2}\simeq2000\,\mathrm{km\,s^{-1}}$,
which is above the escape velocity from the Galaxy \citep{ken+08}.

These order of magnitude estimates approximate a process that depends
on many parameters (the internal orbital parameters of the binary,
and those of the binary's center of mass relative to the MBH). A full
statistical characterization of the outcomes requires 3-body simulations
\citep[e.g.][]{ken+08,zha+10}. However, since $Q\gg1$, it is possible
to apply a simpler approximate treatment (the restricted 3-body problem),
which reduces the number of parameters and yields some general results
\citep{sar+10,kob+12}: The ejection energy is not a strong function
of the penetration depth (0.1\textendash 0.2 of binaries actually
survive deep penetration); The more massive of the two stars carries
a larger fraction of $\Delta\epsilon$, and so if $\epsilon>0$ (unbound),
the heavier member is preferentially ejected, while if $\epsilon<0$,
it is captured (in the limit $\epsilon\to0$ the ejection / capture
probabilities are independent of the stellar mass).

This simple picture of binary separation is further complicated by
the finite size of the stars, which for some initial parameters results
in stellar collisions and occasional mergers instead of ejections
\citet{gin+07,ant+10,ant+11}. A variation on the Hills mechanism
are 4-body interactions between an incoming triple system and the
MBH, which ejects a hyper-velocity short period binary. A binary merger
resulting in a rejuvenated star may explain some of the ``too young''
HVSs observed \citep{per09}.

While the physics of the binary separation process are understood,
quantitative predictions of the ejection rate and the HVS velocity
distribution depend on poorly constrained properties of binaries in
the Galactic Center (binary fraction, mass function, period distribution,
Sec. \ref{ss:BinColl}) and on details of the loss-cone dynamics (Sec.
\ref{s:LC}).

\paragraph*{Observed hyper velocity stars}

At present, only HVSs ejected from the Milky Way are close enough
to be detected. The first HVS was discovered by chance in a radial
velocity survey of blue horizontal branch halo stars. HVS1, a $\sim3\,\Mo$
main sequence B-type star at a distance of $\sim100\,\mathrm{kpc}$
and with velocity $>673\,\mathrm{km}\,\mathrm{s}^{-1}$ relative to
the Galactic MBH, has at least twice the escape velocity from the
Galaxy \citep{bro+05}. Since then more than $20$ unbound B-type
stars were discovered in a systematic radial velocity survey (see
detailed review by \citet{bro15} for a summary of the current status
of observations, observing strategies, other categories of detected
HVSs, null results, and sample contamination by unrelated fast-moving
stars). Almost all the observed properties of the sample (metallicity,
stellar rotation, stellar age and flight time, proper motion, bound
/ unbound ratio, number and production rate) are either consistent
with the Hill's mechanism or not yet constraining enough. Two properties
are in tension with theoretical predictions. The observed velocity
distribution of HVSs does not extend to velocities as high as predicted
\citep{ros+14}, a possible hint that the binary population is softer
(has longer periods) than assumed. The spatial distribution of the
HVS sample shows an anisotropy in Galactic longitude, which is not
expected in the simplest Hill's scenario. A sample of ${\cal O}(100)$
HVSs is required to discriminate between the different variants of
MBH-assisted HVS ejections \citep{ses+07}.

The total HVS ejection rate (including all stellar types) that is
derived from the observed sample, after correcting for sky coverage,
HVS lifespans, and the number fraction of B stars in the mass function,
is $\sim10^{-4}\ \mathrm{yr^{-1}}$. This is close to the full loss
cone case assumed by \citep{hil88}, which can be justified if massive
perturbers, likely GMCs, are efficiently scattering binaries into
the loss-cone (\citealp{per+07}, Sec. \ref{sss:MPs}). 

The trajectories of HVSs, their velocity distribution, and the ratios
between outgoing and infalling HVSs, all probe the Galactic potential,
its symmetries and the distribution of dark matter \citep{yuq+07,ken+08,per+08b,gia+16,ros+16}.
The numbers of detected HVSs are still not high enough for statistically
robust conclusions, but this will improve with the upcoming GAIA astrometric
data \citep{ros+16}. 

\paragraph*{Tidally captured stars }

Tidal binary separation is a leading explanation for the S-star cluster
observed in the central $1^{\prime\prime}\simeq0.04\,\mathrm{pc}$
around $\SgrA$ \citep{gou+03}. This would imply a direct correspondence
between the S-stars and HVSs, and indeed, the number and lifespan
of the S-stars are consistent with the observed HVSs (see above).

The near isothermal distribution of the S-stars eccentricities, $n(e)\mathrm{d}e=2e\mathrm{d}e$
\citep{gil+16} is inconsistent with the high capture eccentricity
of tidal separation, and neither is it consistent with the low eccentricities
that are expected if the S-stars migrated from the observed stellar
disk (Sec. \ref{ss:SgrA}, Sec. \ref{s:disk}). It is necessary to
invoke post-capture/migration dynamical evolution to explain the observations.
$N$-body simulations indicate that fast SRR evolution in a steep
cusp can reproduce the observed S-star eccentricities if they are
tidally captured and in steady state. A dark cusp is necessary for
generating strong enough RR torques. The disk origin scenario is disfavored.
\citep{per+09b}. 

The fainter $B$-stars on the $1\,\mathrm{pc}$ scale, which are on
eccentric orbits and therefore probably not associated with the stellar
disk, may have also been captured there by the Hill's mechanism, and
should then be associated with slower, bound HVSs \citep{mad+14} 

\subsubsection{``Billiard balls'' exchange}

\label{sss:exchange}

The masses of the S-stars around $\SgrA$ are ${\cal O}(10\,\Mo)$,
which is also the mass scale of stellar BHs that are believed to be
strongly concentration around $\SgrA$ due to mass segregation (Sec.
\ref{sss:SScusp}). The similarity in mass scale suggests a possible
connection, which may be realized dynamically by a 3-body exchange
interaction between an incoming B star on a radial orbit, and a stellar
BH on a tight orbit around the MBH. The exchange cross-section is
most effective when the exchanged stars have a similar mass \citep{heg+96}.
Detailed calculations of the exchange cross-section indicate that
$\sim0.25$ of the observed S-stars can be explained this way \citep{ale+04},
the limiting factors being the small exchange cross-section and the
number of B stars available for scattering toward the central arcsec
around $\SgrA$. The continual replacement of mass-segregated stellar
BHs by NS progenitors, and the analogous process of exchange of NSs
by WD progenitors, may ``downsize'' the compact object population
close to the MBH, and play a role in regulating the buildup of the
dark cusp.

\section{Relativistic interactions with a MBH}

\label{s:GRMBH}

The nature of the relativistic interactions accessible for stars near
an MBH depends on $r_{t}/\rbh$, the tidal radius to event horizon
ratio. $\sim1\,\Mo$ MS stars are disrupted outside $\rbh$ for the
$\Mbh<10^{8}\,\Mo$ MBHs considered here (Sec. \ref{p:TidalField}).
At $r\gtrsim r_{t}$, tidal interactions interfere with geodetic motion.
It follows that only weak post-Newtonian effects are expected to be
detectable in the orbits of MS stars \citep[but see][]{fre03}. However,
stellar BHs (and to lesser extent NSs and WDs) can reach the event
horizon unperturbed, and therefore probe strong gravity.

\subsection{Gravitational waves from extreme mass ratio inspirals}

\label{ss:EMRIs}

The emission of GWs from EMRIs provides clean tests of strong gravity.
GR precession decouples the last stages of GW inspiral from the RR
torques of the background stars (Sec. \ref{sss:LCSteadyState}), and
the non-geodetic drag by an accretion disk is negligible \citep{nar00,lev07}.
In addition, EMRIs can provide information about MBH demographics,
stellar BHs, stellar dynamics in galactic nuclei and cosmological
parameters\footnote{The measurables from an EMRI are $\Mbh(1+z)$, $\chi_{\bullet}$ (MBH
dimensionless spin parameter), $\Ms(1+z)$, $e$, $i$ (inclination
angle), $D_{L}$ (luminosity distance) and sky position \citep{bar+04}.
A redshift determination from an electro-magnetic counterpart, if
available, breaks the redshift/mass degeneracy and constrains a combination
of the cosmological parameters $H_{0},\Omega_{M},\Omega_{\Lambda}$.} \citep[See review by][]{ama+07}. The large mass ratio between the
MBH and a stellar BH ($Q\gtrsim10^{5}$) makes them test masses in
the MBH's space-time (once the BH's small effect on space-time, its
``self-force'', is taken into account. \citealp[See review by][]{bar09}). 

GWs are detected by measuring the relative change in distance, the
strain $h=\Delta R/R$, due to the tidal effect of a time-varying
metric. The lowest order contribution to the time-dependent tidal
far-field, $g^{\prime}\sim\Delta g/R$, where $g$ is the gravitational
acceleration, comes from the 4th time-derivative of the moment of
inertia $I\sim MR^{2}$, $g^{\prime}\sim G\ddddot{I}/c^{4}D$, where
$D$ is the distance to the MBH. The measured GW strain is therefore
$h\sim\Delta R/R\sim\int\mathrm{d}t\left(\int\mathrm{dt}g^{\prime}\right)\sim G\ddot{I}/c^{4}D\sim GM/c^{4}RD$,
where the time derivative is expressed by the typical angular frequency
$\Omega^{2}=GM/R^{3}$. For a BH, $R\sim r_{g}$, the strain is $h\sim r_{g}/D$
and the typical frequency is $f\sim1/(2\pi t_{g})=c^{3}/2\pi G\Mbh$.
For example, two grazing MBHs with total mass $10^{7}\,\Mo$ at a
distance of $1\,\mathrm{Gpc}$ produce a strain of $h\sim{\cal O}(10^{-16})$
and $f\sim{\cal O}(10^{-3}\,\mathrm{Hz})$. For an EMRI ($\Ms\ll\Mbh$),
a detailed derivation yields \citep{tho87},
\begin{equation}
h=\frac{\sqrt{2^{7}\pi^{4/3}/15}}{Q}\frac{r_{g}}{D}\left(t_{g}f\right)^{2/3}\simeq9\times10^{-23}\frac{\Ms}{10\,\Mo}\left(\frac{\Mbh}{10^{6}\,\Mo}\frac{f}{10^{-3}\,\mathrm{Hz}}\right)^{2/3}\left(\frac{D}{\,1\,\mathrm{Gpc}}\right)^{-1}\,.
\end{equation}

Planned low frequency (mHz) space-borne GW detectors \citep{eli+16},
which are optimal for the $\Mbh\sim{\cal O}(10^{6}\,\Mo)$ mass scale,
will observe the entire sky. An EMRI spends a few years, and goes
through $>10^{5}$ cycles, while in the detection band. A simple estimate
of the number of expected GW sources, assuming an EMRI rate of $\Gamma_{i}=10^{-6}\,\mathrm{yr^{-1}\,}\mathrm{gal^{-1}}$
(\textbf{Figure} \ref{f:LossRates}) for $N=10^{8}$ Milky Way-like
galaxies inside $z=1$ over a 1 yr mission, suggests ${\cal O}(100)$
EMRIs simultaneously emitting in the detection band. Because of the
tiny strain and possible source confusion, detection of GW from distant
galaxies must rely on pre-calculated waveform templates and on the
statistics provided by the large number of quasi-periodic GW cycles.
The templates depend on the eccentricity of the final orbits, which
reflect the dynamics leading to the inspiral. GW dissipation of energy
and angular momentum generally drives EMRI orbits to circularization.
However, EMRIs that began on eccentric loss-cone orbits tend to retain
a fairly high eccentricity \citep{hop+05}; EMRIs captured by binary
tidal separation will be circular \citep{mil+05}; EMRIs captured
or formed in the disk will be circular and in the equatorial plane
of the MBH (Sec. \ref{ss:hydro}).

\subsection{Stars and pulsars on relativistic orbits in the Galactic Center}

\label{ss:GRorbits}

Post-Newtonian phenomena that may be observed around $\SgrA$ include
the lowest order effects of gravitational redshift and relativistic
Doppler shift, which are already detectable near periapse passage
with available spectroscopy \citep{zuc+05}. Relativistic precession
of stars on very short period orbits could be detected by the adaptive
optics-assisted IR interferometer GRAVITY \citep{eis+11} and test
the ``no-hair'' theorem \citep{wil08}. RR perturbations by background
stars are a major concern, with the exception of polar orbits relative
to the MBH spin axis $\boldsymbol{\chi}_{\bullet}$ \citep{mer+10},
where the out-of plane Lense-Thirring precession and quadrupole precession
provides AI protection against the stellar perturbations (Sec. \ref{sss:eta}).
It is not clear how many stars exist on very relativistic orbits,
given the short survival time so close to the MBH. High precision
measurement of radio pulsars could accurately map space-time around
the MBH \citep[See review by][]{eat+15}. It may be possible to circumvent
the stellar perturbations by using data only from the fraction of
the orbit near periapse, where the perturbations are small \citep{psa+16}.
It should be noted however that only one pulsar (a magnetar, \citealt{ken+13})
has been found to date near $\SgrA$. It is not clear whether this
is due to strong electron scattering along some lines of sight, or
that contrary to expectations, radio emitting NSs are inherently rare
in that environment. Finally, it is possible that EMRIs from very
low-mass stars, which are dense enough to withstand the MBH tides,
will be detected \citep{fre03}, and that gravitational lensing of
background stars by the MBH and stellar black holes (BHs) around it
will be observed  \citep{war+92,ale+99a,ale+01c,ale01,nus+04}.

\section{Star-star collisions near a MBH}

\label{s:sscoll}

The rate of physical star-star collisions, per star, is $\Gamma_{c}\simeq16\sqrt{\pi}\ns\sigma\Rs^{2}\left[1+(\vs/\sigma)^{2}\right]$,
where the term $(\vs/\sigma)^{2}$ expresses the effect of gravitational
focusing, and a Maxwellian velocity distribution with dispersion $\sigma^{2}\sim G\Mbh/r$
is assumed \citep[e.g.][]{bin+08}. Close enough to the MBH, where
$\sigma>\vs$, inside the collision radius $r_{\mathrm{coll}}=Q\Rs=(c/\vs)^{2}r_{g}\sim{\cal O}(10^{5})r_{g}$
(Sec. \ref{p:HighVel}), the collision rate rapidly increases as $\Gamma_{c}\propto r^{-\alpha-1/2}$
(in an $\ns\propto r^{-a}$ cusp) until the mean time between collisions
becomes shorter than the stellar lifespan, and the system is dominated
by physical collisions. Since the kinetic energy in the colliding
star exceeds the binding energy, the collisions lead to stellar destruction
rather than to mergers, whether in a single head-on collision, or
in several grazing collisions that lead to mass loss. Such high velocity
collisions are unique to the near environment of a MBH.

\subsection{Collisional destruction and mergers }

\label{ss:2StarColl}

\paragraph*{Collisional stripping of red giant envelopes }

The absence of brightest red giants ($K<13.5^{\text{m}}$) in the
central $\sim1^{\prime\prime}$ ($\sim0.04$ pc) of the Galactic Center
has led to suggestions that this may be the result of collisional
envelope stripping. The envelope's dynamical and thermal timescales
are short; a hole ``punched out'' by a star passing through it would
be quickly filled. Lasting damage to the giant can only be achieved
if the impactor kicks the giant's core out of the envelope, or forms
a common-envelope binary \citep{liv+88}. This requires a nearly head-on
collision for the high velocities near $\SgrA$, or a collision with
a binary \citep{bai+99}. Giant envelope stripping by collisions with
single stars and binaries is consistent with the observations in the
central $1^{\prime\prime}$, if a high density power-law cusp is assumed
\citep{ale99}. However, it cannot explain the depletion of the higher-luminosity
giants ($K<12^{\text{m}}$) farther out on the $5^{\prime\prime}$
scale. To do that would require a large population of stellar BHs
\citet{dal+09}, which is inconsistent with the drain limit (Sec.
\ref{sss:SScusp}), or with the dynamical mass measurements in the
Galactic Center. 

\paragraph*{Tidal spin-up }

Non-destructive close stellar collisions excite tides on the two stars,
which lead to energy dissipation, tidal torquing and possible mass
loss. The random orbital orientations of successive close encounters
over the lifetimes of long-lived low-mass stars can build up a substantial
stellar spin by random walk (assuming inefficient magnetic breaking).
Over time the low-mass stars around the MBH are expected spin at $0.1$\textendash $0.3$
of the centrifugal breakup velocity \citep{ale+01a}. Such high rotation
may explain the strong coronal radiation that is observed around $\SgrA$
\citep{saz+12}.

\paragraph*{Collisional destructions / mergers}

The region depleted of red giants in the Galactic Center happens to
overlaps with the region where many young blue stars are observed.
This coincidence motivates the search for a ``rejuvenation'' mechanism
\citep{ghe+03}, that can transform an old red giant to a young-looking
stellar object (Sec. \ref{ss:hydro}). Stellar collisions and mergers
were discussed in this context. However, hydrodynamical simulation
overall confirm the simple physical analysis presented above: fast
off-center collisions are inefficient in substantially changing the
star (although many repeated collisions may whittle it gradually down,
\citealt{rau99}), whereas fast head-on collisions are disruptive
\citep{lai+93,fre+05}. While some exotic merger products may exist
around $\SgrA$, it does not seem this can explain a substantial fraction
of the population.

\texttt{}

\subsection{Collisions involving binaries}

\label{ss:BinColl}

Binaries near a MBH are not as dynamically important as binaries in
clusters, where they control the evolution \citep[e.g.][]{bin+08},
since the single stars are tightly bound to the MBH. However, they
may play other roles, for example, in HVS ejection and tidal captures;
tidally captured stars can become EMRIs, or be tidally disrupted;
binaries can acquire a compact companion by evolution or exchange,
and appear as X-ray sources or produce millisecond pulsars that can
be used to probe space-time around the MBH (Sec. \ref{ss:GRorbits});
binary mergers may produce massive blue stars that are observed in
the Galactic Center, whose origin is uncertain. 

The fraction of binaries among the stars inside $r_{h}$ is not well
know. The small number of massive young binaries detected in the Galactic
Center is consistent with the binary fraction in young stellar clusters
elsewhere in the Galaxy \citep{pfu+14}. The over-abundance of transient
X-ray sources in the inner 1 pc of the Galactic Center, identified
as NS/BH low or high-mass X-ray binaries \citep{mun+05} likewise
hints that binaries, or at least high-mass binaries, do exist inside
$r_{h}$. 

Long-lived low-mass binaries are gradually evaporated by 3-body interactions
with single stars in a high density stellar cusp, and their fraction
drops with proximity to the MBH \citep{hop09}. Conversely, the existence
of low-mass binaries near $\SgrA$, if such are detected, will place
an upper bound on the local stellar density (and a lower bound on
the local NR time), and thereby reveal a dark cusp around the MBH
\citep{ale+14}.

Internal collisions and mergers between the binary components can
be induced by Kozai-Lidov oscillations due to the MBH (acting as an
external perturber in the binary's frame, \citealp[See review by][]{nao16}\texttt{)}
\citep{ant+10,pro+15,ste+16}, and also in the course of a binary
separation event (Sec. \ref{sss:TidalSeparation}).

\section{Stars and circumnuclear accretion disks}

\label{s:disk}

A circumnuclear accretion disk is embedded in a dense circumnuclear
cluster. It is plausible that the two are strongly coupled. Many studies
have explored a wide range of possible mechanisms and implications.
These are categorized here in several broad themes.

\subsection{Hydrodynamical star/disk interactions}

\label{ss:hydro}

\subsubsection*{Stellar trapping, growth and destruction}

Stars whose eccentric orbits intersect the disk, experience drag and
gradually settle into circular co-rotating orbits \citep{ost83,sye+91}.
Once embedded in the disk, they migrate inward, in some cases opening
a gap in the disk, and in others continuously growing by accreting
mass from it. The star perturbs the disk and can excite periodic AGN
variability. If the stellar accretion and evolutionary timescales
are faster than the migration, the stars will explode as SNe while
in the disk , thereby raising its metallicity early on, as is observed
in high-$z$ quasars \citep{art+93}. Stars that open a gap in the
disk and migrate to its inner edge before exploding, will interrupt
the gas supply and cause a transient dimming and reddening of the
quasar \citep{goo+04} and ultimately be tidally disrupted. Remnants
trapped in the disk or formed following disk growth and SNe explosions
can reach short-period relativistic orbits and become EMRI GW sources
\citep{lev07}. 

A major puzzle is the suppression of red giants  in the inner $\sim0.5$
pc of the Galactic Center, contrary to theoretical expectations of
a high density cusp (Sec. \ref{ss:SgrA}). Intriguingly, this length-scale
coincides with the extent of the $\sim6\,\mathrm{Myr}$ old stellar
disk, which is believed to have formed from a gravitationally unstable
gas disk. Interactions between such a clumpy gas disk and red giants
that cross it can gradually strip the giants of their envelopes even
over the short time the disk exists at this marginally stable phase,
provided that the giants expand in response to partial stripping \citep{ama+14}.
Hydrodynamical simulations provide some support for this scenario
\citep{kie+16}. Such selective stripping, if true, implies that the
giants do not trace the faint population.

\subsubsection*{Stars and mass accretion rate}

Accretion requires a mechanism to transport angular momentum outward
through the disk, to some sink. Stars can play various roles in the
disk's angular momentum transfer. The gravitational drag on stars
crossing the disk extracts angular momentum from the disk on the NR
timescale \citep{ost83}. This is too slow compared to the viscous
rate typically assumed for AGN disks, unless the the MBH is very low
mass (IMBH scale), or the disk is very thin (cold). SNe in the disk
can redistribute angular momentum at a rate equivalent to that assumed
for a typical AGN accretion disk \citep{roz+95}. In analogy to migration
in proto-planetary disks, a single massive star undergoing type III
(outward) migration in an AGN disk can extract enough angular momentum
to to generate AGN accretion rates \citep{mck+11}.

The accretion rate may be regulated by the rate $\Gamma_{i}(\sigma)$
(Eq. \ref{e:InspiralRate}) at which stars are scattered to eccentric
disk-crossing orbits, and eventually get trapped in it. This connects
$\sigma$ and $\Mbh$, and may explain the $\Mbh/\sigma$ relation
\citep{mir+05}.

\subsection{Gravitational star/disk interactions}

\label{ss:RRwarp}

\subsubsection*{Disk warping}

A thin accretion disk, whose radial inflow velocity is much slower
than the near-Keplerian circular velocity, can be viewed as a set
of concentric mass rings that are coupled by their internal viscous
stresses. VRR exerts torques on the disk (Sec. \ref{ss:RR}) with
spatial coherence scale $\Delta\log a\sim1$ (footnote \ref{fn:RRrcorr})
and temporal coherence timescale $T_{c}^{SQ}$. The VRR torques warp
the disk (assumed to extend over $\log(a_{\max}/a_{\min})\gg1$),
but are countered by the disk's out-of-plane viscous torques that
flatten it. Scaling arguments and simulations \citep{bre+09,bre+12}
show that the gravitational RR torques are stronger than the hydrodynamical
drag torques by stars crossing the disk (Sec. \ref{ss:hydro}) for
$\Mbh\gtrsim\mathrm{few}\times10^{4}\,\Mo$, and that the external
RR torques are stronger than the internal viscous torques for $\Mbh<{\cal O}(10^{7}\,\Mo$).
Therefore disks around lighter MBHs warp under the influence of VRR,
while disks around more massive MBHs rotate as solid bodies. 

Warping by RR affects the physical properties and dynamics of the
disk \citep{bre+12}. The warps expose the disk to the ionizing continuum
from the central source, thereby increasing its viscosity and modifying
its line emission. The continual warping of the disk extracts angular
momentum and thereby increases the mass accretion rate \citep{lod+06}.
The Bardeen-Petterson frame-dragging effect couples the stellar torques
to the MBH spin via the warps. This leads to a jitter in the spin
direction, which may translate to a jitter in the radio jet direction,
if it is aligned with MBH spin axis. 

This RR-driven disk warping scenario can be tested by modeling observed
warps in maser disks. Circumnuclear maser disks are typically found
around MBHs with $\Mbh\sim10^{7}\,\Mo$ and display warps of a few
degrees \citep{kuo+11}. In particular, the well studied maser disk
in AGN NGC~4258 shows an $8^{\circ}$ warp \citep{miy+95}, which
is consistently and naturally reproduced (in the statistical sense)
by stellar RR torques \citep{bre+12}. 

\subsection{Disk fragmentation and star formation}

\label{ss:SFdisk}

The young stellar disk around $\SgrA$ \citep{bar+09,luj+09} provides
compelling indications that star formation can occur \emph{in-situ}
in a circumnuclear disk. This unusual formation mode is quite unlike
that observed in the galactic field, where stars form in dense, cold
self-gravitating molecular clouds. The observed stellar disk is naturally
explained by the gravitational instability and fragmentation of a
gas disk, similar to observed circumnuclear maser disks \citep{mil+04}.
The initial stellar mass function in a fragmenting disk is likely
top-heavy \citep{lev07}. Low mass ($\lesssim3\,\Mo$) young stellar
objects produce copious X-rays, which are not observed. This is consistent
with a top-heavy initial mass function with a $\sim3\,\Mo$ lower-mass
cutoff \citep{nay+05a}.

\section{Future prospects}

\label{s:future}

Over the next two decades, major new instruments are expected to provide
tools and opportunities for studying galactic nuclei, MBHs and strong
gravity. Already operating are the VLT/GRAVITY interferometer \citep{eis+11},
which is looking for low-mass stars on relativistic orbits around
$\SgrA$ with unprecedented astrometric precision, and large scale
transient source surveys (Pan-STARRS \citep{kai+02}, PTF/ZTF \citep{law+09,bel14},
ASAS-SN \citep{sha+14}), which search for TDEs.

Pulsars on relativistic orbits around $\SgrA$ may be detected in
a few years with the next generation millimeter observatories (LMT,
phased-NOEMA, phased-ALMA) and by the radio Square Kilometer Array
\citep{eat+15}. In a decade, extremely large telescopes (E-ELT, TMT)
will dramatically improve the ability to observe fine details in galactic
nuclei. In two decades, the low-frequency space-borne GW detector
eLISA will detect EMRIs to $z\lesssim0.5$ and essentially all MBH\textendash MBH
mergers \citep{eli+16}.

The prospects of precision GR tests with GWs, MBH astronomy with GWs
and TDEs, and detailed observations of the Galactic Center and of
external galactic nuclei with extremely large telescopes, are continuing
to motivate deeper understanding and high-fidelity modeling of these
processes across all the relevant theoretical sub-fields. 

\section*{DISCLOSURE STATEMENT}

The author is not aware of any affiliations, memberships, funding,
or financial holdings that might be perceived as affecting the objectivity
of this review.

\section*{ACKNOWLEDGMENTS}

I am grateful for helpful comments by Ben Bar-Or and Elena Rossi.
This work was supported by the I-CORE Program of the PBC and the ISF
(grant No 1829/12).

\bibliographystyle{ar-style2}
\bibliography{ARAA}

\end{document}